\documentclass{emulateapj}

\newcommand\kms{\ifmmode{\rm km\thinspace s^{-1}}\else km\thinspace s$^{-1}$\fi}
\newcommand\lk{LkCa\,3}

\shortauthors{Torres et al.}
\shorttitle{\lk}

\begin{document}

\submitted{Accepted for publication in the Astrophysical Journal}

\title{The quadruple pre-main sequence system \lk: implications \\ for
  stellar evolution models}

\author{
Guillermo Torres\altaffilmark{1},
Dary Ru\'\i z-Rodr\'\i guez\altaffilmark{2},
Mariona Badenas\altaffilmark{3},
L.\ Prato\altaffilmark{2},
G.\ H.\ Schaefer\altaffilmark{4},
Lawrence H.\ Wasserman\altaffilmark{2},
Robert D.\ Mathieu\altaffilmark{5}, and
David W.\ Latham\altaffilmark{1}
}

\altaffiltext{1}{Harvard-Smithsonian Center for Astrophysics, 60
Garden St., Cambridge, MA 02138, USA; e-mail: gtorres@cfa.harvard.edu}

\altaffiltext{2}{Lowell Observatory, 1400 West Mars Hill Road,
Flagstaff, AZ 86001, USA}

\altaffiltext{3}{Department of Astronomy, Yale University, New Haven,
CT 06520, USA}

\altaffiltext{4}{The CHARA Array of Georgia State University, Mount
Wilson Observatory, Mount Wilson, CA 91023, USA}

\altaffiltext{5}{Department of Astronomy, University of
Wisconsin-Madison, Madison, WI 53706, USA}

\begin{abstract} 
We report the discovery that the pre-main sequence object \lk\ in the
Taurus-Auriga star-forming region is a hierarchical quadruple system
of M stars. It was previously known to be a close ($\sim$0\farcs5)
visual pair, with one component being a moderately eccentric 12.94-day
single-lined spectroscopic binary.  A re-analysis of archival optical
spectra complemented with new near-infrared spectroscopy shows both
visual components to be double-lined, the second one having a period
of 4.06 days and a circular orbit. In addition to the orbital
elements, we determine optical and near-infrared flux ratios,
effective temperatures, and projected rotational velocities for all
four stars. Using existing photometric monitoring observations of the
system that had previously revealed the rotational period of the
primary in the longer-period binary, we detect also the rotational
signal of the primary in the 4.06-day binary, which is synchronized
with the orbital motion.  With only the assumption of coevality, a
comparison of all of these constraints with current stellar evolution
models from the Dartmouth series points to an age of 1.4 Myr and a
distance of 133 pc, consistent with previous estimates for the region
and suggesting the system is on the near side of the Taurus complex.
Similar comparisons of the properties of \lk\ and of the well-known
quadruple pre-main sequence system GG\,Tau with the widely used models
from the Lyon series for a mixing length parameter of $\alpha_{\rm ML}
= 1.0$ strongly favor the Dartmouth models.
\end{abstract}


\section{Introduction}
\label{sec:introduction}

The study of multiplicity among pre-main sequence (PMS) stars is of
key importance to our understanding of the process of star formation,
and to interpret the observed multiplicities of normal stars in the
solar neighborhood, which are evolved from the younger populations. A
recent review of stellar multiplicity by \cite{Duchene:13} noted that
while imaging surveys of PMS stars by many teams have turned up
significant numbers of visual binaries, our knowledge of tighter
systems typically discovered in spectroscopic surveys is considerably
more incomplete among young stars.  This is unfortunate, as close
binaries for which the orbital elements are known provide very useful
constraints on star formation theory from the distributions of their
mass ratios \citep[e.g.,][]{Bate:09}, as well as their periods and
other properties. They also present an opportunity to improve our
understanding of tidal circularization theory \citep{Zahn:89,
Mathieu:92, Melo:01, Witte:02, Meibom:05, Mazeh:08}. Importantly, in
favorable cases they allow the determination of the dynamical masses
of the components, which are essential for testing models of stellar
evolution for young stars \citep[see, e.g.,][]{Hillenbrand:04,
Mathieu:07, Simon:08}. Relatively few PMS stars have this information
available, and as a result evolution models for young stars are much
more poorly constrained than those for their main-sequence
counterparts.

An account of our state of knowledge about multiple systems among PMS
stars by \cite{Mathieu:94} listed a grand total of 25 spectroscopic
binaries for which orbits had been established, of which half were
unpublished at the time. Since then relatively few long-term
spectroscopic surveys have been carried out \citep[e.g.,][]{Covino:01,
Melo:03, Prato:07, Guenther:07, Nguyen:12}, and while some have
produced new systems with orbital solutions, most additions to the
list of binaries with known orbits have come from serendipitous
discoveries.  Slightly more than half of the known PMS spectroscopic
binaries are double-lined, which are especially valuable in that they
provide mass ratios. In recent years significant efforts have been
carried out with good success to detect the faint secondaries of some
of the single-lined systems by observing in the near infrared (NIR),
where the contrast is more favorable \citep[e.g.,][]{Mazeh:02,
Prato:02a, Prato:02b, Mace:12, Simon:13}.

\lk\ (also V1098\,Tau, 2MASS\,J04144797+2752346, HBC\,368) was listed
by \cite{Mathieu:94} as a single-lined spectroscopic binary in the
Taurus-Auriga star-forming region with an orbital period of 12.941
days and an eccentricity of 0.20, although the particulars of the
orbital solution were not reported there. It was also known at the
time to be a visual binary with an angular separation of 0\farcs47
\citep{Leinert:93}, implying a hierarchical configuration of at least
three stars assuming the visual components are physically associated.
Its PMS status was established earlier by \cite{Herbig:86} on the
basis of its strong \ion{Ca}{2} H and K emission and the presence of
the \ion{Li}{1} $\lambda$6707 line in absorption.  Its spectral type
is most often listed as M1, and from the strength of the H$\alpha$
emission line \citep[$\sim$2.5\,\AA\ equivalent width;][]{Hartmann:87}
it is considered a weak-line T Tauri star (WTTS).  No infrared excess
or sub-millimeter emission has been observed \citep[e.g.,][]{Simon:95,
Dutrey:96, Stassun:01, Andrews:05}, consistent with the absence of any
appreciable surrounding disk material.

Beyond their initial use by \cite{Mathieu:94} to report the discovery
of \lk\ as a single-lined spectroscopic binary, the original spectra
of this object have not been exploited to the fullest extent now
possible with the significant improvements in spectroscopic analysis
techniques that have taken place in the intervening years. This has
prompted us to revisit the system. The motivation for this paper is
the discovery that \lk\ is in fact a hierarchical quadruple system, in
which each of the visual components is a double-lined spectroscopic
binary. Full details of the orbital solutions are presented, which
incorporate not only the archival spectroscopic material gathered by
\cite{Mathieu:94} but also near-infrared spectroscopy obtained more
recently that spatially resolves the visual pair. \lk\ thus joins the
small group of double-lined PMS systems with well-established orbits,
contributing two new entries to the list.

The quadruple nature of the system enables much stronger tests of PMS
models than would otherwise be possible. We take advantage of this
opportunity here.  Our spectroscopic observations, along with
additional high-resolution imaging we have acquired to better
characterize the visual binary, allow us to infer the key physical
properties of the four stars in the system by comparison with current
models of stellar evolution, including an independent estimate of the
distance.

\section{Observations}
\label{sec:observations}

\subsection{Optical spectroscopy}
\label{sec:optical}

The visible-light spectroscopy of \lk\ was collected at the
Harvard-Smithsonian Center for Astrophysics (CfA) over a period of a
little more than nine years beginning in November of 1985. This is
largely the same material used by \cite{Mathieu:94} to obtain the
preliminary single-lined orbit of the system, but the details are
reported here for the first time.  Observations were acquired with
nearly identical echelle spectrographs \citep[Digital
Speedometers;][]{Latham:85, Latham:92} attached to the 1.5\,m
Tillinghast reflector at the F.\ L.\ Whipple Observatory on Mount
Hopkins (AZ), and the 4.5\,m-equivalent MMT also on Mount Hopkins,
prior to its conversion to a monolithic 6.5\,m telescope. A single
echelle order 45\,\AA\ wide was recorded using intensified
photon-counting Reticon detectors, at a central wavelength of about
5190\,\AA\ that includes the lines of the \ion{Mg}{1}\,b triplet. The
resolving power of these instruments was $\lambda/\Delta\lambda
\approx 35,000$, and the signal-to-noise ratios of the observations
range from 9 to 21 per resolution element of 8.5\,\kms. A total of 58
exposures were obtained, although four were later discarded because of
moonlight contamination or other problems.

\begin{figure}
\epsscale{1.15}
\plotone{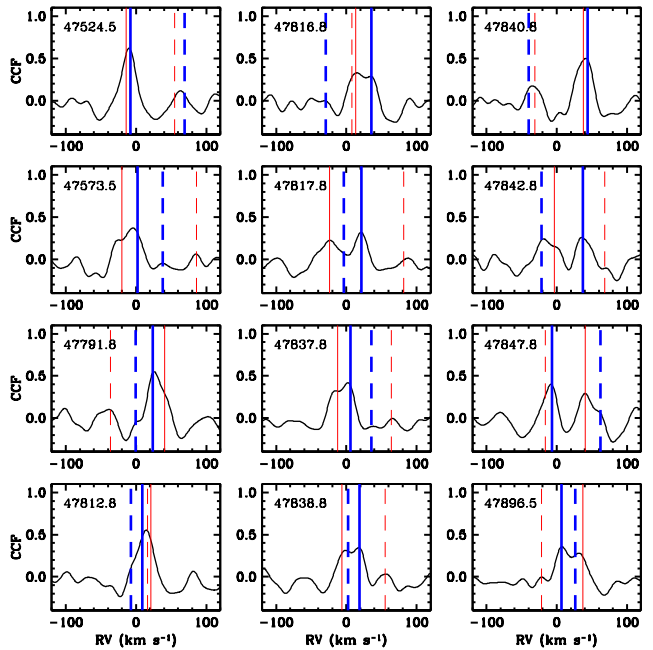}
\figcaption[]{Sample one-dimensional cross-correlation functions for
\lk\ from our optical CfA spectra, with the final RVs of the four
components at the corresponding phases as measured with QUADCOR marked
with vertical lines. Thicker lines (blue in the online color version)
correspond to \lk\,A (solid for the primary, dashed for the
secondary), and thin lines (red) are for \lk\,B. The Julian date of
the observation (${\rm HJD}-2,\!400,\!000$) is indicated in each
panel. Individual spectra have signal-to-noise ratios too low to
permit a clear visual identification of spectral lines from the
different components, and are thus not shown.\label{fig:ccf}}
\end{figure}

The standard, one-dimensional cross-correlation functions for these
spectra clearly show a peak corresponding to the star reported by
\cite{Mathieu:94} to have a variable radial velocity (RV) with a
period of 12.94 days. Occasionally, however, a second peak of similar
height is also visible (see Figure~\ref{fig:ccf}). Shortly after the
\cite{Mathieu:94} publication the spectra of \lk\ were subjected to an
analysis with TODCOR, a two-dimensional cross-correlation technique
that had just been introduced by \cite{Zucker:94}. TODCOR is designed
to measure radial velocities from double-lined spectra, and uses two
templates, one for each set of lines. It was discovered that the
velocities from the second set of lines were not changing with the
same period as the first set, but varied instead with a shorter period
of 4.06 days.  Given that \lk\ is a near equal brightness visual
binary with an angular separation ($\sim$0\farcs5) that is smaller
than the width of the spectrograph slit (1\arcsec), this immediately
suggested the quadruple nature of the system, with each visual
component being in turn a spectroscopic binary.  We refer to the
12.94-day binary as system A, with components Aa and Ab, and to the
stars in the other system as Ba and Bb, although from these
observations alone we cannot identify which visual component is A and
which is B.

This discovery languished for a time, and only more recently did we
return to the system in an attempt to detect the secondary star of the
12.94-day system. For this we reanalyzed the spectra using TRICOR, an
extension of TODCOR to three dimensions \citep{Zucker:95}. The effort
was successful, and showed the faint secondary to be only half as
massive as the primary. A similar exercise with TRICOR resulted in the
detection of the secondary of the 4.06-day binary, which also has
about half the mass of its primary.

\begin{deluxetable*}{lcrrrrcrrrr}
\tablecaption{Heliocentric radial velocities for \lk\ from CfA\label{tab:cfarvs}}
\tablehead{
\colhead{HJD} &
\colhead{Phase} &
\colhead{$RV_{\rm Aa}$} &
\colhead{$\sigma_{\rm Aa}$} &
\colhead{$RV_{\rm Ab}$} &
\colhead{$\sigma_{\rm Ab}$} &
\colhead{Phase} &
\colhead{$RV_{\rm Ba}$} &
\colhead{$\sigma_{\rm Ba}$} &
\colhead{$RV_{\rm Bb}$} &
\colhead{$\sigma_{\rm Bb}$}
\\
\colhead{($2,\!400,\!000+$)} &
\colhead{(\lk\,A)} &
\colhead{(\kms)} &
\colhead{(\kms)} &
\colhead{(\kms)} &
\colhead{(\kms)} &
\colhead{(\lk\,B)} &
\colhead{(\kms)} &
\colhead{(\kms)} &
\colhead{(\kms)} &
\colhead{(\kms)}
}
\startdata
    47524.6317  & 0.4974 & $-$7.98 &  1.66 &   69.23 &  6.51  &  0.4221    & $-$13.76 &  2.67 &  55.05 &  11.84 \\
    47543.6560  & 0.9674 & 46.80 &  1.85 &  $-$46.59 &  7.23  &  0.0991    &  44.02 &  2.96 & $-$32.18 &  13.15 \\
    47544.5958  & 0.0400 & 41.91 &  1.82 &  $-$46.16 &  7.11  &  0.3302    &  $-$6.40 &  2.91 &  55.87 &  12.94 \\
    47545.5693  & 0.1152 & 30.84 &  1.98 &   $-$8.94 &  7.76  &  0.5695    & $-$18.93 &  3.18 &  75.34 &  14.13 \\
    47555.6846  & 0.8968 & 37.86 &  1.85 &  $-$38.64 &  7.23  &  0.0563    &  47.38 &  2.96 & $-$59.42 &  13.15 \\
    47573.6815  & 0.2874 &  1.97 &  2.06 &   37.78 &  8.08  &    0.4807    & $-$20.05 &  3.31 &  86.39 &  14.70 \\
    47791.8706  & 0.1466 & 23.83 &  1.60 &   $-$0.48 &  6.26  &  0.1213    &  40.60 &  2.56 & $-$36.38 &  11.39 \\
    47810.9024  & 0.6171 & $-$2.93 &  2.06 &   47.35 &  8.08  &  0.8002    &  28.11 &  3.31 &  19.40 &  14.70 \\
    47811.9441  & 0.6976 &  7.41 &  1.98 &   34.83 &  7.76  &    0.0563    &  49.65 &  3.18 & $-$63.19 &  14.13 \\
    47816.8581  & 0.0773 & 35.64 &  1.82 &  $-$29.37 &  7.11  &  0.2643    &  13.35 &  2.91 &   7.77 &  12.94 \\
    47817.8777  & 0.1561 & 21.58 &  1.91 &   $-$3.51 &  7.48  &  0.5150    & $-$23.91 &  3.06 &  81.84 &  13.61 \\
    47818.7831  & 0.2260 & 13.75 &  1.85 &   30.64 &  7.23  &    0.7376    &  11.54 &  2.96 &   5.53 &  13.15 \\
    47822.7730  & 0.5343 & $-$7.55 &  1.62 &   55.88 &  6.34  &  0.7185    &  14.18 &  2.60 &   6.91 &  11.54 \\
    47837.7335  & 0.6903 &  5.99 &  1.85 &   35.93 &  7.23  &    0.3964    & $-$12.60 &  2.96 &  64.12 &  13.15 \\
    47838.7451  & 0.7685 & 18.77 &  1.66 &    2.40 &  6.51  &    0.6451    &  $-$6.36 &  2.67 &  55.15 &  11.84 \\
    47839.7495  & 0.8461 & 35.90 &  1.85 &  $-$13.02 &  7.23  &  0.8921    &  36.47 &  2.96 & $-$29.94 &  13.15 \\
    47840.7298  & 0.9218 & 44.13 &  2.02 &  $-$39.68 &  7.92  &  0.1331    &  37.99 &  3.24 & $-$31.38 &  14.41 \\
    47842.7792  & 0.0802 & 37.31 &  1.88 &  $-$21.38 &  7.35  &  0.6369    &  $-$3.40 &  3.01 &  68.21 &  13.38 \\
    47845.7562  & 0.3102 &  3.17 &  1.40 &   50.46 &  5.49  &    0.3688    &  $-$7.08 &  2.25 &  61.82 &   9.99 \\
    47846.8502  & 0.3947 & $-$7.85 &  1.79 &   55.85 &  7.00  &  0.6377    &  $-$7.78 &  2.87 &  66.30 &  12.73 \\
    47847.8078  & 0.4687 & $-$6.86 &  1.79 &   62.22 &  7.00  &  0.8732    &  40.86 &  2.87 & $-$16.04 &  12.73 \\
    47868.8798  & 0.0969 & 34.59 &  2.45 &  $-$17.70 &  9.60  &  0.0536    &  46.14 &  3.93 & $-$72.48 &  17.47 \\
    47869.7600  & 0.1650 & 22.56 &  1.34 &   13.18 &  5.24  &    0.2700    &   8.14 &  2.15 &   1.26 &   9.54 \\
    47870.8110  & 0.2462 &  5.48 &  1.12 &   20.35 &  4.40  &    0.5284    & $-$19.69 &  1.80 &  98.67 &   8.00 \\
    47871.7089  & 0.3155 & $-$4.41 &  1.39 &   43.28 &  5.44  &  0.7491    &  14.47 &  2.23 &   6.69 &   9.90 \\
    47873.7834  & 0.4758 & $-$9.62 &  1.71 &   60.79 &  6.69  &  0.2591    &  11.34 &  2.74 &   4.99 &  12.18 \\
    47878.7773  & 0.8617 & 36.01 &  1.49 &  $-$12.20 &  5.84  &  0.4868    & $-$26.55 &  2.39 &  86.12 &  10.62 \\
    47896.5666  & 0.2363 &  6.64 &  1.58 &   26.22 &  6.18  &    0.8602    &  37.41 &  2.53 & $-$21.89 &  11.25 \\
    47897.5831  & 0.3148 & $-$0.39 &  1.60 &   50.57 &  6.26  &  0.1101    &  42.12 &  2.56 & $-$34.62 &  11.39 \\
    47898.5768  & 0.3916 & $-$4.76 &  1.56 &   49.87 &  6.11  &  0.3544    &  $-$6.53 &  2.50 &  63.72 &  11.12 \\
    47900.5812  & 0.5465 & $-$5.66 &  1.56 &   55.83 &  6.11  &  0.8472    &  36.94 &  2.50 & $-$28.18 &  11.12 \\
    47901.5750  & 0.6233 & $-$2.31 &  1.62 &   46.30 &  6.34  &  0.0915    &  45.42 &  2.60 & $-$43.68 &  11.54 \\
    47902.6324  & 0.7050 &  9.49 &  1.64 &   13.53 &  6.42  &    0.3515    &  $-$7.67 &  2.63 &  59.96 &  11.69 \\
    47903.6486  & 0.7835 & 20.08 &  1.52 &   $-$3.59 &  5.97  &  0.6013    & $-$15.97 &  2.44 &  58.75 &  10.86 \\
    47904.6876  & 0.8638 & 35.19 &  1.51 &  $-$17.34 &  5.90  &  0.8567    &  38.25 &  2.42 & $-$32.03 &  10.74 \\
    47905.7074  & 0.9426 & 44.86 &  1.56 &  $-$43.52 &  6.11  &  0.1075    &  41.80 &  2.50 & $-$36.67 &  11.12 \\
    47906.7332  & 0.0218 & 46.15 &  1.52 &  $-$45.76 &  5.97  &  0.3596    &  $-$9.23 &  2.44 &  51.82 &  10.86 \\
    47908.8453  & 0.1850 & 18.71 &  1.88 &   13.93 &  7.35  &    0.8789    &  39.42 &  3.01 & $-$30.94 &  13.38 \\
    47958.6627  & 0.0343 & 43.85 &  1.51 &  $-$37.93 &  5.90  &  0.1262    &  39.91 &  2.42 & $-$39.07 &  10.74 \\
    47960.6862  & 0.1907 & 14.65 &  1.54 &   17.75 &  6.04  &    0.6237    &  $-$8.38 &  2.47 &  67.04 &  10.99 \\
    47965.6562  & 0.5747 & $-$5.97 &  1.52 &   51.71 &  5.97  &  0.8455    &  36.34 &  2.44 & $-$34.40 &  10.86 \\
    47994.6117  & 0.8121 & 23.37 &  1.66 &   $-$0.56 &  6.51  &  0.9641    &  40.93 &  2.67 & $-$45.64 &  11.84 \\
    47995.6062  & 0.8889 & 40.56 &  2.21 &  $-$39.45 &  8.64  &  0.2086    &  30.21 &  3.54 & $-$17.69 &  15.72 \\
    47996.6027  & 0.9659 & 45.75 &  1.91 &  $-$49.83 &  7.48  &  0.4536    & $-$19.22 &  3.06 & 101.16 &  13.61 \\
    48167.9722  & 0.2074 & 12.62 &  1.56 &   23.07 &  6.11  &    0.5838    & $-$16.34 &  2.50 &  58.90 &  11.12 \\
    48175.9425  & 0.8233 & 27.87 &  1.07 &   $-$9.54 &  4.17  &  0.5433    & $-$22.26 &  1.71 &  85.27 &   7.59 \\
    48194.7904  & 0.2796 &  4.58 &  1.23 &   43.45 &  4.80  &    0.1769    &  30.45 &  1.97 &  $-$9.18 &   8.74 \\
    48195.7994  & 0.3576 & $-$3.94 &  1.26 &   52.34 &  4.95  &  0.4250    & $-$14.42 &  2.03 &  68.47 &   9.00 \\
    48200.8740  & 0.7497 & 13.47 &  1.51 &   18.43 &  5.90  &    0.6725    &  $-$1.12 &  2.42 &  68.04 &  10.74 \\
    48201.8223  & 0.8230 & 28.73 &  1.71 &  $-$17.04 &  6.69  &  0.9057    &  42.57 &  2.74 & $-$35.80 &  12.18 \\
    48291.7312  & 0.7701 & 18.87 &  1.46 &    7.43 &  5.71  &    0.0093    &  49.69 &  2.34 & $-$72.09 &  10.40 \\
    48943.8887  & 0.1614 & 19.25 &  1.68 &    2.27 &  6.60  &    0.3386    &  $-$7.05 &  2.70 &  60.77 &  12.01 \\
    49268.9754  & 0.2804 &  0.85 &  1.85 &   51.55 &  7.23  &    0.2594    &  17.34 &  2.96 &  10.57 &  13.15 \\
    49705.7008  & 0.0256 & 41.91 &  1.68 &  $-$50.84 &  6.60  &  0.6260    & $-$10.34 &  2.70 &  55.63 &  12.01 
\enddata
\end{deluxetable*}

The TRICOR velocities obtained from these two analyses are likely
biased to some extent due to line blending from the fourth star, whose
presence is ignored in each case.  With proof that the spectra are in
fact quadruple-lined, we proceeded to reanalyze them one more time
with QUADCOR, which is an extension of TODCOR to four dimensions
\citep{Torres:07}. In previous studies of solar-type stars with
similar spectroscopic material we have typically used synthetic
spectra as templates for the cross-correlations, based on model
atmospheres by R.\ L.\ Kurucz. Here, however, all components are M
stars, and calculated spectra become less realistic at these low
temperatures, particularly at high resolution. Instead, we used strong
exposures of M stars obtained with the same instrumentation, covering
a range of spectral types. We applied rotational broadening following
the prescription of \cite{Gray:05}. A large number of template
combinations was tried, guided initially by the results described
below from the near-infrared spectra. In those observations the visual
components are spatially resolved, so the spectra are only
double-lined, making template selection easier. For the rotational
broadening we were also guided by the results from \cite{Nguyen:12},
described later in Sect.~\ref{sec:orbits}, who resolved some of the
components spectroscopically with observations of much higher
signal-to-noise ratio than ours.

For the 12.94-day binary, the templates that provided the best match
to the observations, as measured by the cross-correlation value
averaged over all exposures, are GJ\,49 (M1.5; adopted ${\rm RV} =
-5.48$\,\kms) and GJ\,699 (M4; ${\rm RV} = -108.77$\,\kms),
rotationally broadened to 12\,\kms\ and 4\,\kms, respectively. For the
4.06-day binary we used GJ\,846 (M0; ${\rm RV} = +18.61$\,\kms) and
GJ\,15\,A (M2; ${\rm RV} = +12.72$\,\kms), broadened to 16\,\kms\ and
14\,\kms. Our final QUADCOR velocities are presented in
Table~\ref{tab:cfarvs}, along with the individual uncertainties. These
are typically 1.7 and 5.7\,\kms\ for stars Aa and Ab, and 2.6 and
11\,\kms\ for stars Ba and Bb.  To monitor the zero-point of our
velocity system we obtained exposures of the dusk and dawn sky, and
applied small run-to-run corrections as described by \cite{Latham:92}.
These corrections are included in measurements listed in
Table~\ref{tab:cfarvs}.

In addition to the radial velocities we determined the light ratios
between the stars at the mean wavelength of our spectra
\citep[5190\,\AA; see][]{Torres:07}. The values, averaged over all
exposures, are $\ell_{\rm Ab}/\ell_{\rm Aa} = 0.161 \pm 0.012$,
$\ell_{\rm Ba}/\ell_{\rm Aa} = 0.938 \pm 0.042$, and $\ell_{\rm
Bb}/\ell_{\rm Aa} = 0.223 \pm 0.018$.  The fractional light
contributions are thus 0.43 (Aa), 0.07 (Ab), 0.40 (Ba), and 0.10 (Bb).
As a result the combined light of A is the same as that of B, at least
at these wavelengths.

\subsection{Near-infrared spectroscopy}
\label{sec:infrared}

Near-infrared spectra were obtained at the Keck\,II 10\,m telescope
(Mauna Kea, HI) using the facility, cross-dispersed cryogenic
spectrograph NIRSPEC \citep{McLean:98, McLean:00}.  NIRSPEC employs a
1024$\times$1024 InSb array detector, which can accommodate several
high-resolution spectral orders simultaneously.  We observed in the
$H$ band centered at $\sim$1.555\,$\mu$m, corresponding to NIRSPEC
order 49.  The advantage of this spectral region is that while it
lacks significant contamination from telluric absorption lines, thus
avoiding the need to divide by the spectrum of an early-type star,
numerous OH night sky emission lines are distributed across the
spectral range \citep{Rousselot:00}, supplying an excellent means of
inherent wavelength and zero-point calibration.  It is also rich in
deep atomic and molecular lines found in G- through M-type stars.
Spectra were recorded in sets of four exposures in an `ABBA' dither
pattern to allow for background subtraction and elimination of bad
pixels.  A two-pixel slit width yielded a spectral resolution of
$\lambda/\Delta\lambda \approx 30,\!000$. In total, five observations
were obtained in a period between 2003 and 2010
(Table~\ref{tab:lowellrvs}).

On UT 2006 December 14 and 2010 December 12 we used NIRSPEC in
high-angular resolution mode behind the Keck\,II adaptive optics (AO)
system.  This facilitated separation of the $\sim$0\farcs5 pair,
\lk\,A and B.  Natural seeing conditions on these nights were
$\sim$1\arcsec\ and 0\farcs5, respectively.  Because of multiple
reflections in the complex AO light path, integration times per
exposure were 300\,s even for these relatively bright stars.  The
NIRSPEC slit is re-imaged behind the AO system by a factor of
$\sim$11, making it prohibitively small for collecting light from OH
night sky emission lines.  Comparison lamp lines were thus recorded to
accomplish the dispersion solution and set the zero-point.

On the other three nights (UT 2003 November 5, 2004 December 25, and
2010 November 22) we did not use AO.  Although the seeing was
relatively good (0\farcs4, 0\farcs3, and 0\farcs6, respectively) the
point spread functions (PSFs) of the spectral traces overlapped in the
two-dimensional spectra.  With the higher throughput, shorter
integration times of 60 to 120\,s per exposure were adequate.

\begin{figure}
\epsscale{1.15}
\plotone{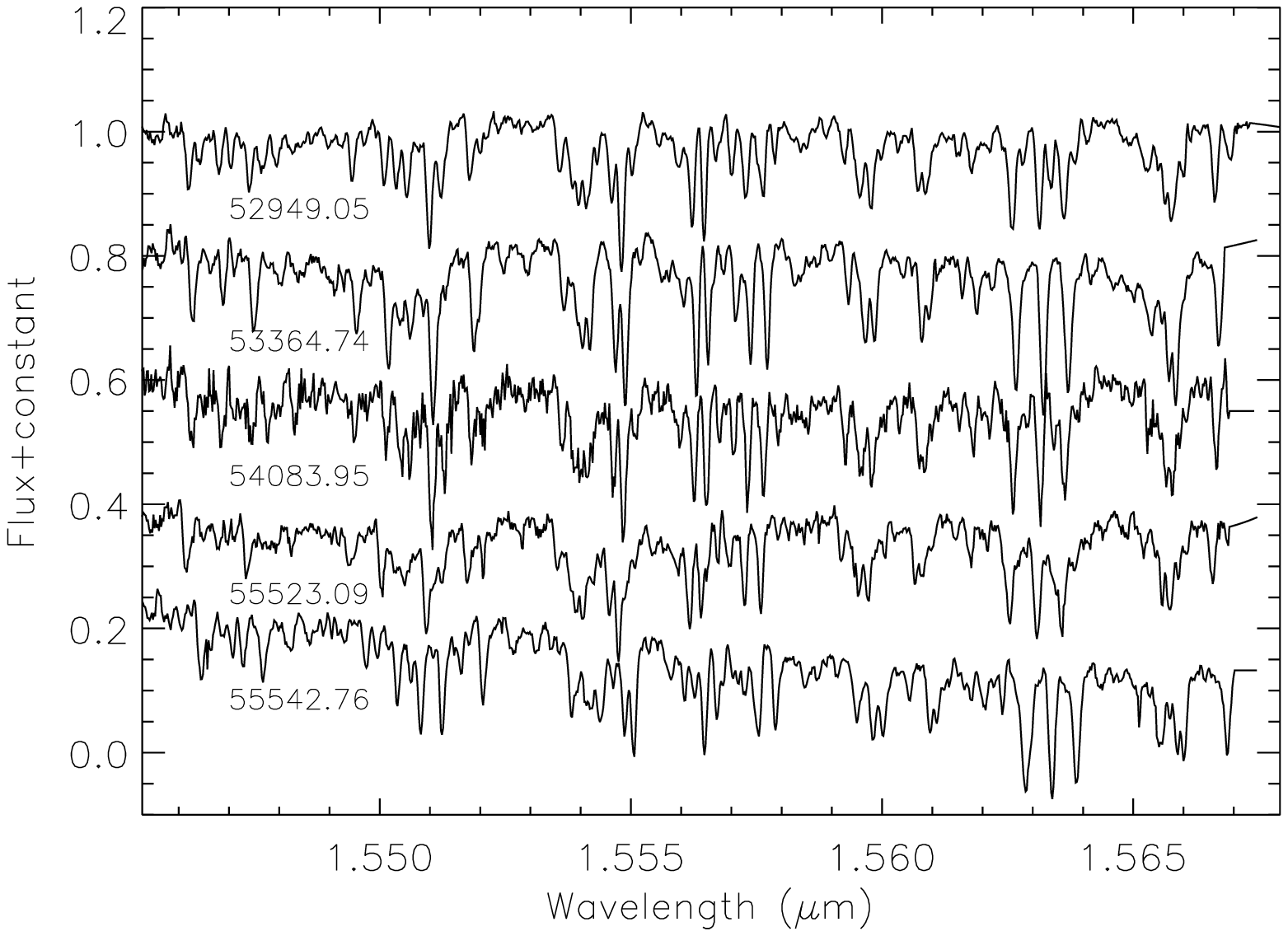}
\figcaption[]{Five epochs of NIR Keck spectra for \lk\,A (West) in
NIRSPEC order 49, with barycentric corrections applied; Julian dates
of the observations (HJD$-$2,400,000) are indicated. The spectra have
been normalized to unity and an arbitrary additive constant used to
offset the spectra for display.\label{fig:IRa}}
\end{figure}

\begin{figure}
\epsscale{1.15}
\plotone{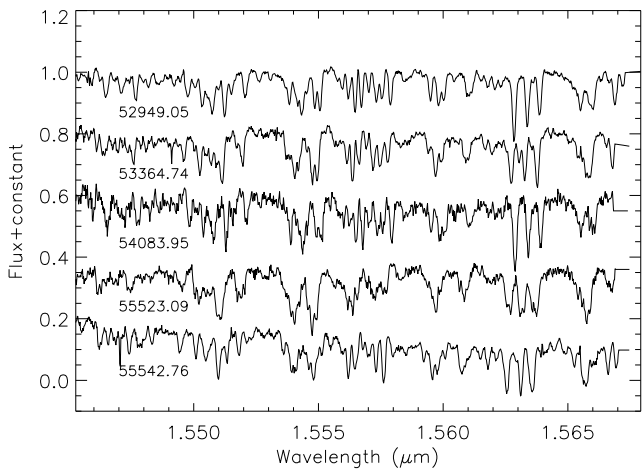}
\figcaption{Same as Figure~\ref{fig:IRa}, for \lk\,B
(East).\label{fig:IRb}}
\end{figure}

Spectra were extracted with the REDSPEC IDL package\footnote{
\url{http://www2.keck.hawaii.edu/inst/nirspec/redspec/index.html}},
available on the Keck website and specifically designed for the
analysis of NIRSPEC data.  These procedures correct for spatial and
spectral distortions in the two-dimensional spectra.  Reduction of the
AO data was trivial; the same procedure was followed as in, e.g.,
\cite{Rosero:11}.  However, to extract individual spectra from the
overlapping PSFs in the non-AO data it was necessary to use REDSPEC to
extract the rectified two-dimensional spectra and then apply further
operations to separate \lk\,A and B.  Customized IDL procedures were
employed to determine the best Gaussian parameters for the A and B
PSFs.  The stellar signal for each of the 1024 columns in the
cross-dispersion direction of the rectified two-dimensional spectrum
was then fit with two overlapping model Gaussians, varying the heights
to identify the flux (the Gaussian maximum) of each component
individually.  The reduced and barycenter-corrected spectra for order
49 are shown in Figures~\ref{fig:IRa} and \ref{fig:IRb}.


\begin{deluxetable*}{lcrrrrcrrrr}
\tablecaption{Heliocentric radial velocities for \lk\ in the near infrared\label{tab:lowellrvs}}
\tablehead{
\colhead{HJD} &
\colhead{Phase} &
\colhead{$RV_{\rm Aa}$} &
\colhead{$\sigma_{\rm Aa}$} &
\colhead{$RV_{\rm Ab}$} &
\colhead{$\sigma_{\rm Ab}$} &
\colhead{Phase} &
\colhead{$RV_{\rm Ba}$} &
\colhead{$\sigma_{\rm Ba}$} &
\colhead{$RV_{\rm Bb}$} &
\colhead{$\sigma_{\rm Bb}$}
\\
\colhead{($2,\!400,\!000+$)} &
\colhead{(\lk\,A)} &
\colhead{(\kms)} &
\colhead{(\kms)} &
\colhead{(\kms)} &
\colhead{(\kms)} &
\colhead{(\lk\,B)} &
\colhead{(\kms)} &
\colhead{(\kms)} &
\colhead{(\kms)} &
\colhead{(\kms)}
}
\startdata
    52949.0510  & 0.6347 & $-$0.94 &  0.70 &   40.52 &  1.10  &  0.9858 &  47.79 &  1.50 & $-$54.81 &   2.30 \\
    53364.7389  & 0.7544 & 15.55 &  0.70 &   10.78  &  1.10  &    0.1804 &  27.35 &  1.50 & $-$16.42 &   2.30 \\
    54083.9544  & 0.3272 & $-$3.14 &  0.70 &   46.05 &  1.10  &  0.9956 &  48.67 &  1.50 & $-$54.49 &   2.30 \\
    55523.0858  & 0.5269 & $-$6.78 &  0.70 &   52.86 &  1.10  &  0.7982 &  19.09 &  1.50 &  $-$9.52 &   2.30 \\
    55542.7560  & 0.0467 & 40.81 &  0.70 &  $-$41.28 &  1.10  &  0.6340 &  $-$9.45 &  1.50 &  55.44 &   2.30
\enddata
\end{deluxetable*}

Since our procedures separate \lk\,A and B, the NIR spectra are only
double-lined, and the velocity analysis is considerably simpler than
in the optical. Furthermore, with previous knowledge of the orbits
from our optical spectra, the NIR observations allow us to
unambiguously identify star A ($P = 12.94$ days) as the Western
component, and star B ($P = 4.06$ days) as the Eastern one.

Radial velocities were measured using the TODCOR algorithm of
\cite{Zucker:94}. As with our optical spectra, we correlated each of
our observations against a grid of observed template spectra of high
signal-to-noise ratio \citep{Prato:02a} seeking the highest
correlation coefficient averaged over all exposures. These IR
template spectra were rotationally broadened by convolving them with
the line-broadening function of \cite{Gray:05}. For \lk\,A the best
fit was produced combining the M2 star GJ\,752\,A (adopted ${\rm RV} =
+34.2\,\kms$) rotationally broadened to 12\,\kms\ with the M4 star
GJ\,402 (${\rm RV} = -3.1\,\kms$) spun up to 6\,\kms, as the primary
and secondary templates, respectively. For \lk\,B the best templates
were 61\,Cyg\,B (K7; ${\rm RV} = -65.4\,\kms$) and GJ\,436 (M2.5;
${\rm RV} = +7.8\,\kms$), rotated to 18 and 14\,\kms,
respectively. The nominal spectral types of these templates are within
one sub-type of the templates adopted in the optical for the same
components, and the rotational velocities are also quite similar
(within 2\,\kms). Average values for the latter are reported later in
Sect.~\ref{sec:properties}.

The flux ratios in the $H$ band were found to be $0.47 \pm 0.06$ for
\lk\,A (ratio Ab/Aa), and $0.52 \pm 0.08$ for \lk\,B (Bb/Ba), which
are averages over the 5 observations in each case. The NIR radial
velocities in the heliocentric frame are presented in
Table~\ref{tab:lowellrvs}, along with their estimated uncertainties.

\subsection{High-resolution imaging}
\label{sec:astrometry}

We observed \lk\ on UT 2008 January 17 with the adaptive optics system
at the W.\ M.\ Keck Observatory, using the near-infrared camera NIRC2
\citep{Wizinowich:00}. A series of 10 dithered images were collected
in the narrow-band $H$ and $K$ continuum regions, using a 5-point
dither pattern with an offset of 2\arcsec. Each image consisted of 10
co-added 0.5\,s exposures. The images were flat-fielded using
dark-subtracted, medianed dome flats. Pairs of dithered images were
subtracted to remove the sky background.

We used the brighter, Western component (\lk\,A) as a PSF reference to
measure the relative separation and flux ratio of the Eastern
component relative to the Western one. We corrected the positions for
geometric distortions in the detector, used a plate scale of $9.952
\pm 0.001$ mas~pixel$^{-1}$, and subtracted $0\fdg252 \pm 0\fdg009$
from the raw position angles to correct for errors in the orientation
of the camera relative to the true North \citep{Yelda:10}. Taking the
average of the $H$- and $K$-band images, we measured an angular
separation of $483.73 \pm 0.28$ mas and a position angle of $69\fdg068
\pm 0\fdg035$ East of North, at Julian date 2,454,482.872. We also
determined the flux ratio of the Eastern relative to Western component
to be $0.942 \pm 0.011$ at $H$, and $0.9404 \pm 0.0074$ at $K$.
Uncertainties were derived by analyzing multiple images individually
and computing the standard deviation.

While the angular separation between the visual components of \lk\ has
remained the same since its discovery by \cite{Leinert:93}, the
position angle has decreased significantly by almost 10\arcdeg. This
could be a sign of orbital motion, but it may also be the result of a
small difference in proper motion between two unbound members of the
star-forming region. Our spectroscopic orbital solutions described in
the next section indicate that the center-of-mass velocities of \lk\,A
and B are very close to each other (within 0.4\,\kms), but this, too,
is consistent with either scenario.  Though the current observations
are insufficient to confirm the physical association, an estimate of
the likelihood of a chance alignment in Taurus strongly favors the
binary hypothesis. Based on the stellar densities near \lk\ reported
by \cite{Gomez:93}, we compute a probability of finding two unbound
stars within only 0\farcs5 of each other to be $\sim10^{-7}$, and for
the remainder of the paper we will assume that \lk\,A and B are
physically bound. At the distance to the object of 133\,pc estimated
below in Sect.~\ref{sec:rotation}, the projected linear separation of
the visual binary is approximately 64\,AU, and the orbital period
roughly 400\,yr.

\section{Spectroscopic orbital solutions}
\label{sec:orbits}

\begin{figure}
\epsscale{1.1}
\plotone{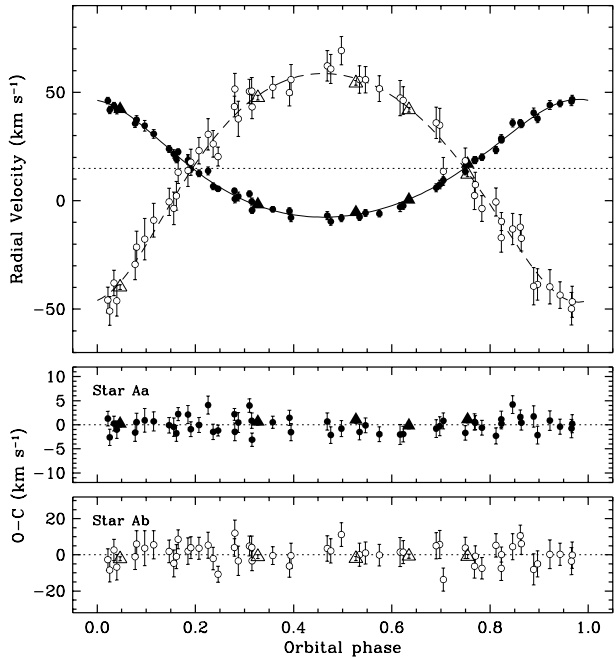}
\figcaption[]{Radial-velocity measurements for \lk\,A (West) along
with our weighted orbital fit. Filled symbols are for star Aa (circles
for the CfA observations, triangles for the NIR measurements), and
open symbols for star Ab. The dotted line in the top panel represents
the center-of-mass velocity of the binary, and phase 0.0 corresponds
to periastron passage. Residuals are shown at the
bottom.\label{fig:orbitA}}
\end{figure}

\begin{figure}
\epsscale{1.1}
\plotone{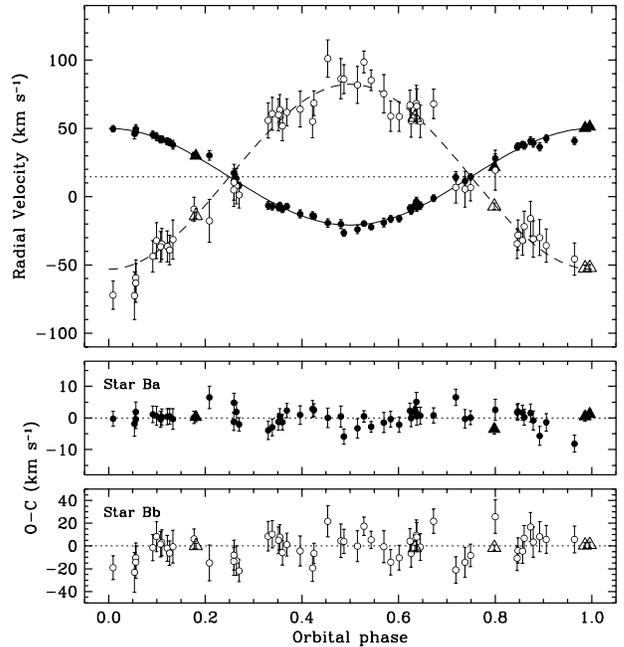}
\figcaption[]{Same as Figure~\ref{fig:orbitA} for \lk\,B (East). The
orbit is assumed to be circular. Phase 0.0 corresponds to the phase of
maximum primary velocity.\label{fig:orbitB}}
\end{figure}

The optical and near-infrared RVs were used simultaneously to compute
weighted orbital solutions for each of the two binaries. A velocity
offset for the NIR data (in the sense CfA minus Keck) was added as a
free parameter to take into account possible differences in the zero
points. We found these offsets to be significant for both
binaries. The resulting orbital elements and other derived quantities
are listed in Table~\ref{tab:orbits}. While the eccentricity of the
\lk\,A orbit is clearly non-zero, that of \lk\,B in our initial
solution was very small ($e = 0.025 \pm 0.011$), and statistically
insignificant according to the test of \cite{Lucy:71}, with a false
alarm probability of 0.08. We therefore chose to adopt here the
parameters from a solution for a circular orbit. As noted earlier, the
velocities of the centers of mass, $\gamma$, are essentially the same
for the two binaries (Table~\ref{tab:orbits}).  Graphical
representations of the observations and final orbit models are shown
in Figure~\ref{fig:orbitA} and Figure~\ref{fig:orbitB}. The NIR
velocities in these figures have been corrected for the offsets
mentioned above.

\begin{deluxetable*}{lcc}
\tablecaption{Spectroscopic orbital solutions for \lk.\label{tab:orbits}}
\tablehead{
\colhead{\hfil~~~~~~~~~~~~~Parameter~~~~~~~~~~~~~~} &
\colhead{\lk\,A (West)} & \colhead{\lk\,B (East)}
}
\startdata
\multispan{3}{Adjusted quantities\hfil} \\
\noalign{\vskip 2pt}
~~~$P$ (days)                          &  12.941865~$\pm$~0.000075\phn     & 4.0676115~$\pm$~0.0000085 \\
~~~$\gamma$ (\kms)\tablenotemark{a}    &  +14.97~$\pm$~0.22\phn\phs        & +14.59~$\pm$~0.35\phn\phs \\
~~~$\Delta RV$ (\kms)\tablenotemark{b} &  +1.46~$\pm$~0.35\phs             & +2.52~$\pm$~0.71\phs \\
~~~$K_{\rm a}$ (\kms)                  &  27.16~$\pm$~0.31\phn             & 35.37~$\pm$~0.45\phn \\
~~~$K_{\rm b}$ (\kms)                  &  52.75~$\pm$~0.77\phn             & 67.7~$\pm$~1.2\phn \\
~~~$e$                                 &  0.1735~$\pm$~0.0090              & 0 (fixed)           \\
~~~$\omega_{\rm a}$ (deg)              & 10.2~$\pm$~2.7\phn                & \nodata          \\
~~~$T$ (${\rm HJD}-2,\!400,\!000$)\tablenotemark{c}     & 48,501.777~$\pm$~0.094\phm{2,222} & 48,499.1416~$\pm$~0.0080\phm{2,222} \\
\noalign{\vskip 2pt}
\multispan{3}{Derived quantities\hfil} \\
\noalign{\vskip 2pt}
~~~$M_{\rm a}\sin^3 i$ ($M_{\sun}$)    &  0.432~$\pm$~0.015                & 0.303~$\pm$~0.013 \\
~~~$M_{\rm b}\sin^3 i$ ($M_{\sun}$)    &  0.2221~$\pm$~0.0062              & 0.1585~$\pm$~0.0050 \\
~~~$q\equiv M_{\rm b}/M_{\rm a}$       &  0.5148~$\pm$~0.0085              & 0.522~$\pm$~0.012 \\
~~~$a_{\rm a}\sin i$ (10$^6$ km)       &  4.760~$\pm$~0.052                & 1.979~$\pm$~0.025 \\
~~~$a_{\rm b}\sin i$ (10$^6$ km)       &  9.25~$\pm$~0.13                  & 3.788~$\pm$~0.069 \\
~~~$a \sin i$ ($R_{\sun}$)             &  20.13~$\pm$~0.21\phn             & 8.29~$\pm$~0.10 \\
\noalign{\vskip 2pt}
\multispan{3}{Other quantities pertaining to the fit\hfil} \\
\noalign{\vskip 2pt}
~~~$N_{\rm obs}$ (CfA / Keck)          & 54 / 5                            & 54 / 5 \\
~~~Time span (yr)                      & 21.95                             & 21.95 \\
~~~$\sigma_{\rm RV}$ for CfA (\kms)         & 1.71 / 5.75                       & 2.64 / 11.48 \\
~~~$\sigma_{\rm RV}$ for Keck (\kms)        & 0.76 / 1.56                       & 1.97 / 1.18
\enddata
\tablenotetext{a}{Formal uncertainties exclude any errors in the absolute RVs of the cross-correlation
templates.}
\tablenotetext{b}{Velocity offset in the sense $\langle$CfA$-$Keck$\rangle$.}
\tablenotetext{c}{For \lk\,A the time listed corresponds to periastron
passage; for \lk\,B it is the time of maximum primary velocity. In
both cases the epoch is the one nearest to the average time of
observation, to minimize the uncertainty.}
\end{deluxetable*}

In a recent study by \cite{Nguyen:12}, a handful of radial-velocity
measurements were reported for \lk\ in the context of a spectroscopic
multiplicity survey in Chamaeleon and Taurus-Auriga. The authors
measured three of the components at each of four epochs, but without
full knowledge of the complexity of the system they were unable to
provide a consistent explanation of the observed RV changes at the
time. Our spectroscopic orbits now permit us to ascertain that their
`star 1' is the one we refer to as Aa, while their stars 2 and 3 are
generally Ba and Bb, respectively, but occasionally another component
or a blend.

\section{Physical properties}
\label{sec:properties}

Under the working assumption that the four stars in \lk\ are coeval,
our estimates of the flux ratios along with other constraints may be
used with a set of PMS evolutionary models to infer the age of the
system, as well as its distance, $D$.  The minimum masses from our
spectroscopic orbits additionally allow us derive the inclination
angles of the two binaries, $i_{\rm A}$ and $i_{\rm B}$. We describe
this in the next section. The observables we incorporate, aside from
the minimum masses, are listed in Table~\ref{tab:constraints}, and
include $V(RI)_{\rm C}JHK_s$ apparent magnitudes from the literature,
flux ratios measured here in the $H$ and $K$ bands as well as in $V$
(converted from 5190\,\AA), and estimates of the effective
temperatures of the four stars. A comparison of our \lk\ B/A flux
ratios from Sect.~\ref{sec:astrometry} with similar measurements from
the literature indicates significant variability in $H$ and $K$, which
is not unexpected for young and active stars such as these. For this
reason we have used a weighted average of all available B/A estimates,
rather than our own single-epoch measurement. The flux ratios in $V$
and $H$ measured between individual stars are already averages of
several epochs from our optical and NIR spectroscopy.

In order to obtain the effective temperatures we made the assumption
that each component of \lk\ is adequately represented by the M stars
adopted as templates, both in the optical QUADCOR analysis and in the
near-infrared TODCOR analyses. Since each component used a different
template in the optical than in the NIR (see Sect.~\ref{sec:optical}
and Sect.~\ref{sec:infrared}), two estimates of $T_{\rm eff}$ may be
obtained for each of the four stars Aa, Ab, Ba, and Bb.  A common
practice in the field of PMS stars is to assign temperatures directly
from spectral types using a look-up table, of which there are
many. Here we have taken a more quantitative approach independent of
the spectral type labels.  When available, we adopted temperature
estimates for the eight template stars derived from NIR spectroscopy
performed by \cite{Rojas-Ayala:12}.  This was the case for GJ\,402,
GJ\,699, 61\,Cyg\,B, GJ\,436, and GJ\,15\,A. In other cases we made
our own $T_{\rm eff}$ estimates from optical and NIR colors in the
literature along with the recent color/temperature calibrations of
\cite{Boyajian:12}.  This procedure was used for GJ\,752\,A, GJ\,49,
and GJ\,846.  For the photometric temperature determinations we
require also an estimate of the reddening, which we obtained from four
different sources as listed in Table~\ref{tab:constraints}. The
adopted average with a conservative uncertainty is $E(B-V) = 0.10 \pm
0.05$, which is in line with other estimates for the region.
Extinction corrections for each passband were applied following
\cite{Cardelli:89}. Finally, we averaged the temperatures for the
optical and NIR templates used for each component of \lk; these mean
values are reported also in Table~\ref{tab:constraints}.

\begin{deluxetable}{lcc}
\tablecaption{Observational constraints for \lk\ for comparison with
stellar evolution models.\label{tab:constraints}}
\tablehead{
\colhead{Parameter} & \colhead{Value} & \colhead{Source}}
\startdata
$V$ (mag)                   &  12.09~$\pm$~0.03\phn  & 1 \\
$R_{\rm C}$ (mag)           &  11.05~$\pm$~0.04\phn  & 1 \\
$I_{\rm C}$ (mag)           &   9.79~$\pm$~0.04      & 1 \\
$J$ (mag)                   &  8.363~$\pm$~0.032     & 2 \\
$H$ (mag)                   &  7.625~$\pm$~0.023     & 2 \\
$K_s$ (mag)                 &  7.423~$\pm$~0.021     & 2 \\
Ab/Aa $V$-band flux ratio   &  0.154~$\pm$~0.012     & 3 \\
Ba/Aa $V$-band flux ratio   &  0.989~$\pm$~0.044     & 3 \\
Bb/Aa $V$-band flux ratio   &  0.232~$\pm$~0.019     & 3 \\
Ab/Aa $H$-band flux ratio   &   0.47~$\pm$~0.06      & 4 \\
Bb/Ba $H$-band flux ratio   &   0.52~$\pm$~0.08      & 4 \\
B/A $H$-band flux ratio     &  0.878~$\pm$~0.079     & 5 \\
B/A $K$-band flux ratio     &  0.891~$\pm$~0.048     & 5 \\
Adopted $E(B-V)$ (mag)      &   0.10~$\pm$~0.05      & 6 \\
$T_{\rm eff}$ of Aa (K)    &  3570~$\pm$~100\phn    & 7 \\
$T_{\rm eff}$ of Ab (K)    &  3290~$\pm$~100\phn    & 7 \\
$T_{\rm eff}$ of Ba (K)    &  3870~$\pm$~150\phn    & 7 \\
$T_{\rm eff}$ of Bb (K)    &  3490~$\pm$~150\phn    & 7 \\
$v \sin i$ of Aa (\kms)    &  12~$\pm$~2\phn        & 8 \\
$v \sin i$ of Ab (\kms)    &   5~$\pm$~4            & 8 \\
$v \sin i$ of Ba (\kms)    &  17~$\pm$~1\phn        & 8 \\
$v \sin i$ of Bb (\kms)    &  14~$\pm$~1\phn        & 8 \\
$P_{\rm rot}$ of star Aa (days) &  7.38~$\pm$~0.40      & 9 \\
$P_{\rm rot}$ of star Ba (days) &  4.06~$\pm$~0.13      & 9
\enddata
\tablecomments{Sources are:
1.\ \cite{Kenyon:95};
2.\ 2MASS \citep{Cutri:03};
3.\ Measured from our optical spectra (mean wavelength of 5190\,\AA)
and converted to the $V$ band;
4.\ Measured from our NIR spectra;
5.\ Weighted average of one measurement from this work
(Sect.~\ref{sec:astrometry}) and several others (two in $H$ and five
in $K$) from \cite{Leinert:93}, \cite{Ghez:93}, \cite{Woitas:01a},
\cite{Woitas:01b}, \cite{White:01}, and \cite{McCabe:06};
6.\ Average of estimates based on the dust maps from
\cite{Hakkila:97}, \cite{Schlegel:98}, \cite{Drimmel:03}, and
\cite{Amores:05};
7.\ Estimates derived here based on spectroscopic determinations for
the template stars from \cite{Rojas-Ayala:12} or based on
color/temperature calibrations by \cite{Boyajian:12};
8.\ Average of determinations based on our optical and NIR spectra,
with conservative uncertainties;
9.\ Measured here using the photometric observations of
  \cite{Norton:07} (see text).}

\end{deluxetable}

\section{Comparison with stellar evolution models}
\label{sec:models}

The collection of observational constraints listed in
Table~\ref{tab:constraints}, along with the minimum masses from
Table~\ref{tab:orbits}, were compared in a $\chi^2$ sense with PMS
isochrones for solar metallicity from the Dartmouth series
\citep{Dotter:08}. Specifically, we explored a large number of
combinations of ages, distances $D$, and orbital inclination angles
$i_{\rm A}$ and $i_{\rm B}$ over wide ranges of these variables and in
small steps. For each trial set of inclination angles and age we
computed the absolute masses of the four stars using our spectroscopic
minimum masses. We also read off their temperatures from the
corresponding isochrone, for comparison with our measured $T_{\rm
eff}$ values, as well as their brightness in the relevant bandpasses,
from which the flux ratios can be computed.  Finally, with the trial
distance and the adopted reddening we predicted the apparent
magnitudes needed to compare with the values in
Table~\ref{tab:constraints}.

\begin{figure}
\epsscale{1.15}
\plotone{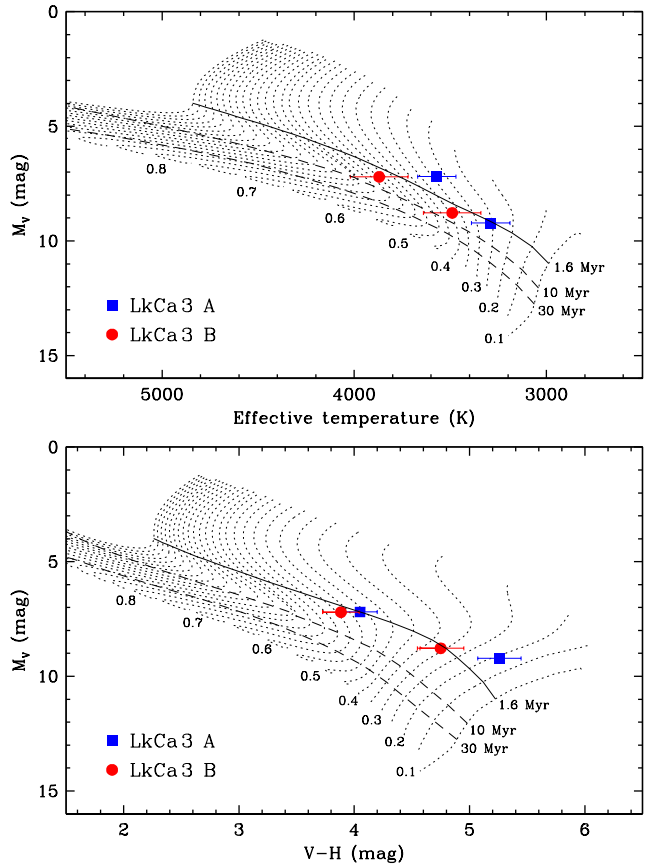}
\figcaption[]{Observations for the four stars in \lk\ compared against
PMS stellar evolution models from the Dartmouth series by
\cite{Dotter:08}. The best-fit 1.6-Myr isochrone is shown with a solid
line (see text).  Evolutionary tracks represented with dotted lines
are labeled with their mass, in solar units; \emph{Top:} Diagram of
absolute visual magnitude versus effective temperature; \emph{Bottom:}
Same as above, illustrating the smaller scatter that results when
using the $V-H$ color instead of $T_{\rm eff}$.\label{fig:models}}
\end{figure}

We applied a simple grid search procedure to find the set of four
variables that yields the best overall fit (lowest $\chi^2$), and
arrived at a satisfactory solution for a distance of $D =
127_{-13}^{+16}$\,pc, an age of $1.6_{-0.5}^{+0.9}$\,Myr, and best-fit
inclination angles for the two binary orbits of $i_{\rm A} =
70.0_{-4.5}^{+5.5}$ deg and $i_{\rm B} = 57.0_{-2.5}^{+3.0}$ deg. The
1-$\sigma$ uncertainties were derived by perturbing all observables in
a Monte Carlo fashion assuming they are normally distributed,
repeating the isochrone fitting 1000 times, and computing the 15.85
and 84.13 percentiles of the resulting distributions. The observations
and best-fit 1.6-Myr isochrone are displayed in a diagram of absolute
visual magnitude versus $T_{\rm eff}$ in the top panel of
Figure~\ref{fig:models}, along with evolutionary tracks from the same
models. The dispersion around the 1.6-Myr isochrone in this plane is
due mostly to errors in the effective temperatures.  Indeed, the lower
panel of the figure shows that the scatter is significantly reduced
when plotting $V-H$ instead of $T_{\rm eff}$, which we are able to do
because this color index can be measured individually for all four
stars from the information given in Table~\ref{tab:constraints}. The
temperature errors may arise in part from the limited selection of
templates we have, both in the optical and in the NIR, from
uncertainties in the calibrations involved in the $T_{\rm eff}$
determinations, and from differences in surface gravity between dwarfs
and PMS stars \citep[see, e.g.,][]{Luhman:99}.

Our age estimate for \lk\ is consistent with other estimates of a few
Myr for T~Tauri stars in Taurus-Auriga \citep[e.g.,][]{Kenyon:95,
Luhman:98, Palla:02, Briceno:02}, but is of course model-dependent.
Similarly, our distance estimate agrees with the canonical value for
the region of 140\,pc.  \lk\ is projected onto the general area of the
Lynds\,1495 molecular cloud at the Western edge of the Taurus-Auriga
star-forming region.  Very precise trigonometric parallax measurements
with the Very Long Baseline Array (VLBA) have shown that different
areas within the complex are at slightly different distances. In
particular, PMS stars near Lynds\,1495 including Hubble\,4,
HDE\,283572, and V773\,Tau (which happens to also be a quadruple
system and is only 20\arcmin\ away from \lk) have an average distance
of about 130\,pc \citep{TorresR:07, TorresR:12}, whereas stars on the
Eastern side of the complex \citep[T~Tau\,N and
HP\,Tau/G2;][]{Loinard:07, TorresR:09} tend to have larger VLBA
distances of 147--160\,pc. Our estimate of $127_{-13}^{+16}$\,pc for
\lk\ is formally more consistent with the lower values than the higher
ones.  There are also differences in the proper motions of these two
groups, and in their radial velocities, with the stars near
Lynds\,1495 having a lower mean RV around +15\,\kms\ compared to +18
or +19\,\kms\ for the far side of the complex. The average RV for \lk\
is +14.8\,\kms\ (see Table~\ref{tab:orbits}), and its proper motion is
also more similar to those of Hubble\,4, HDE\,283572, and V773\,Tau
than to those of T~Tau\,N or HP\,Tau/G2.  Thus, both the kinematic and
distance evidence suggest that \lk\ is on the near side of the
Taurus-Auriga complex.

The best fit to the Dartmouth models indicates the two binaries are
nearly identical in their physical properties. The orbital inclination
angles $i_{\rm A}$ and $i_{\rm B}$ combined with our minimum masses
result in absolute masses for both primary stars (Aa and Ba) of about
0.51\,$M_{\sun}$, and masses for the secondaries (Ab and Bb) that are
both about 0.27\,$M_{\sun}$. The radii predicted from the best-fit
isochrone for these masses are approximately 1.56\,$R_{\sun}$ for the
primaries and 1.22\,$R_{\sun}$ for the secondaries.  The similarity
between the properties of \lk\,A and B is consistent with their
near-equal brightness, as reported earlier in Sect.~\ref{sec:optical}
and Sect.~\ref{sec:astrometry}.  At face value the measurements in
Table~\ref{tab:constraints} suggest star A is slightly brighter than B
in the $H$ band ($\Delta H_{\rm B-A} = +0.14 \pm 0.10$), but fainter
in $V$ ($\Delta V_{\rm B-A} = -0.061 \pm 0.044$)\footnote{These
estimates are averages over time, though not necessarily over the same
time interval for the different passbands.}.  While these differences
may perhaps be real, and would be explained by the fact that the flux
of each visual component is the sum of the contributions of two young
stars, their significance is marginal.

\subsection{Rotation as an additional constraint on models}
\label{sec:rotation}

The orbital periods of \lk\,A (12.94 days) and \lk\,B (4.06 days) are
on either side of the observed tidal circularization period for PMS
stars (considered as a more or less coeval population of young
objects), which represents the transition between circular and
eccentric orbits \citep{Mathieu:92, Mazeh:08}. This transition period
is currently believed to be near 7.6 days, and is marked by the binary
system RX~J1603.9$-$3938, which has the longest-period circular orbit
among bona-fide PMS stars \citep{Melo:01, Covino:01}. Our finding that
the orbit of \lk\,B is essentially circular is consistent with this
picture. Two other effects of tidal forces, the alignment of the spin
axes with the orbital axis and the synchronization of the stellar
rotations with the mean orbital motion, are predicted to occur much
earlier than circularization \citep[e.g.,][]{Hut:81}, so we may
reasonably assume that these conditions also apply to the two
components of \lk\,B. In other words, $i_{\rm rot} = i_{\rm orb}$ and
$P_{\rm rot} = P_{\rm orb}$. If that is the case, then our measurement
of the projected rotational velocities (which are strictly $v \sin
i_{\rm rot}$) provides a way to infer the mean stellar densities, as
described by \cite{Torres:02}, if we assume solid-body rotation. From
the orbital elements and $v \sin i$ the density $\rho$ of star Ba in
solar units may be expressed as
\begin{eqnarray}
\label{eq:densB}
\log\rho_{\rm Ba} &=& -1.8724 - 2\log P_{\rm orb} + 2 \log(K_{\rm Ba}+K_{\rm Bb}) + \nonumber \\
                  & & \log K_{\rm Bb} - 3 \log(v_{\rm Ba}\sin i)
\end{eqnarray}
where $P_{\rm orb}$ is in days, the velocity semi-amplitudes $K_{\rm
Ba}$ and $K_{\rm Bb}$ are in \kms, and the numerical constant is an
update of that of \cite{Torres:02}.  A similar expression holds for
$\log\rho_{\rm Bb}$.

The non-circular orbit we find for \lk\,A (with $e = 0.1735 \pm
0.0090$) is consistent with the fact that its period is longer than
the transition period for PMS stars.  Rotational synchronization is
therefore not guaranteed, so we can not assume that $P_{\rm rot} =
P_{\rm orb}$. We note, however, that long-term photometric monitoring
of \lk\ has shown a clear periodic signal with an amplitude of one or
two tenths of a magnitude attributed to rotational modulation due to
spots, although the multiplicity of the system has previously made it
difficult to attribute the signal to any particular component.  The
period reported by \cite{Bouvier:95} from a 45-day observing campaign
is $7.2 \pm 0.6$ days, and a similar result of 7.38 days was obtained
by \cite{Norton:07} over a two-month period. More extensive
observations by \cite{Grankin:08} between 1992 and 1998 yielded
rotation periods in close agreement for each of 5 separate seasons
having sufficient observations, with an average of 7.35 days. More
recently \cite{Xiao:12} reported a period of 3.689 days from
observations carried out over two months. This is very nearly half of
the previous values, suggesting it may be a harmonic of the true
period.

\begin{figure}[b!]
\epsscale{1.15}
\vskip 10pt
\plotone{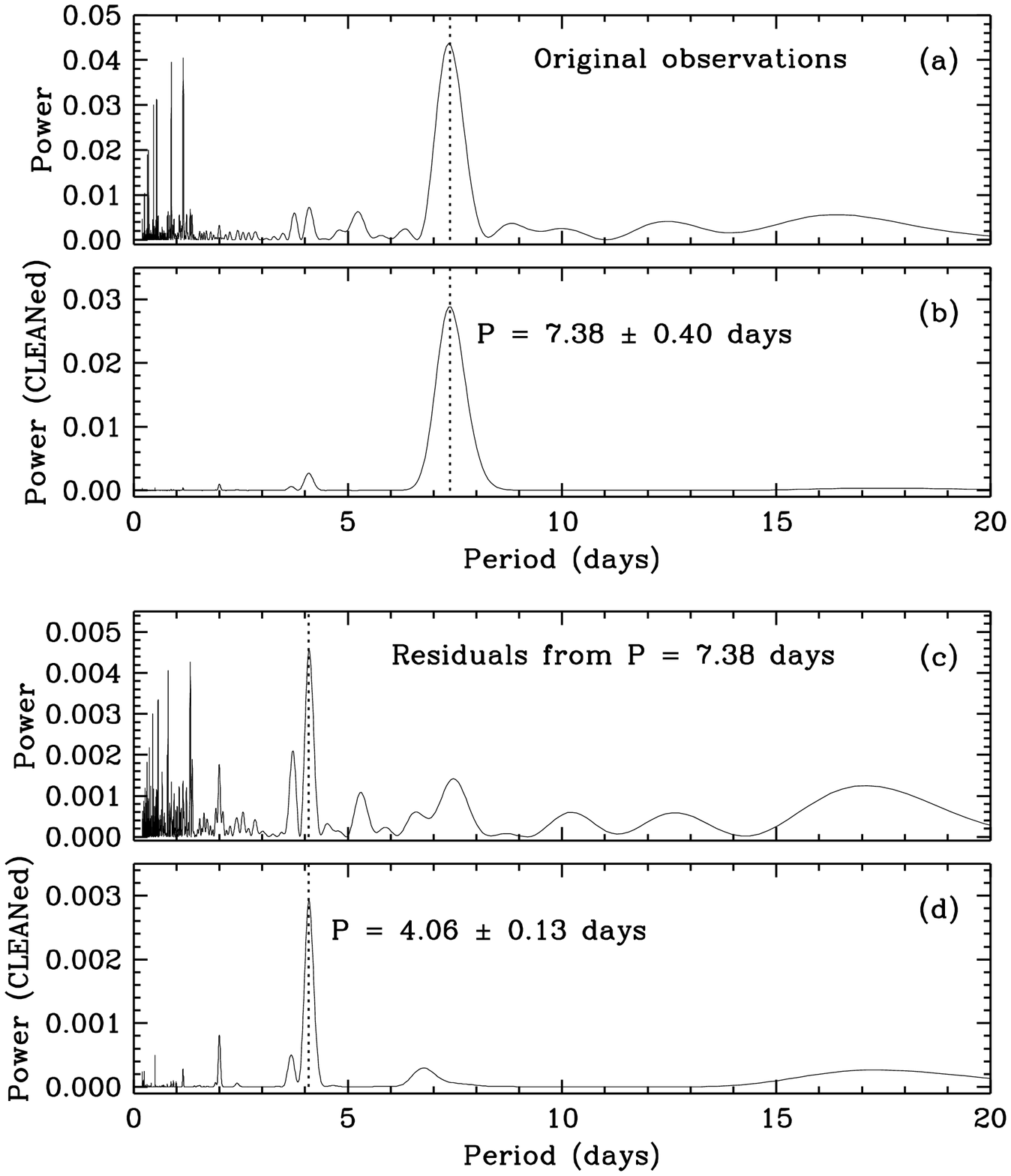}
\figcaption[]{\emph{(a)} Lomb-Scargle periodogram of the photometric
observations of \lk\ by \cite{Norton:07}; \emph{(b)} Same as above,
processed with the CLEAN algorithm of \cite{Roberts:87}, and showing
the 7.38-day rotational signal from star Aa; \emph{(c)} Lomb-Scargle
periodogram of the residuals from the \cite{Norton:07} data, after
subtracting a sine-curve fit to remove the 7.38-day signal; \emph{(d)}
CLEANed power spectrum of the same residuals, revealing the highly
significant 4.06-day signal that we attribute to star
Ba.\label{fig:periodograms}}
\end{figure}

The rotation signal in \lk\ is likely to be due to either star Aa or
Ba, simply on the basis that they are much brighter than the
secondaries. We proceed under this assumption, and return to it
below. Furthermore, if our earlier hypothesis of synchronized rotation
for Ba and Bb is correct, then the signal would be associated with
star Aa, which is also marginally the brighter.  Evidence supporting
this comes from the intensive photometric observations by
\cite{Norton:07}, totaling about 1350 measurements over a two-month
period in a broad passband similar to $V$+$R$. Their power spectrum of
the observations shows a hint of a second peak near a period of 4
days. Our reanalysis of those data is illustrated in
Figure~\ref{fig:periodograms}. The top panel presents a Lomb-Scargle
periodogram of the original observations with the 7.38-day signal
clearly visible. The photometric amplitude of this signal is
approximately 0.10 mag, and we estimate the uncertainty in the period
to be 0.40 days The peaks at periods under two days are all aliases
with the 1-day cycle of the observations.  This is seen more clearly
in the second panel, where we have subjected the same data to the
CLEAN algorithm of \cite{Roberts:87} that reduces the impact of the
window function. We then subtracted a sine curve with a period of 7.38
days from the original measurements, and computed a Lomb-Scargle
periodogram of the residuals (Figure~\ref{fig:periodograms}c).  A
prominent peak is revealed at a period of $4.06 \pm 0.13$ days, with a
false alarm probability smaller than $10^{-3}$, determined from Monte
Carlo simulations. This signal has a smaller photometric amplitude of
about 0.05 mag. Once again the peaks at much shorter periods are
aliases, as shown by the CLEANed power spectrum in the bottom panel of
the figure. The close similarity with the orbital period of \lk\,B
($P_{\rm orb} = 4.0676115$; Table~\ref{tab:orbits}) strongly suggests
that this second signal corresponds to the rotational signature of
star Ba, tidally locked to the mean orbital motion, as we had
assumed. We conclude that the 7.38-day signal corresponds indeed to
star Aa, which is then spinning more rapidly than the expected
pseudo-synchronous rate \citep[$P_{\rm pseudo} = 11.0$
days;][]{Hut:81}, and also faster than the periastron rate (9.0 days).

The evidence that the above signals originate on the primaries of
\lk\,A and B rather than the secondaries relies on statistics from
variability studies of non-accreting periodic T~Tauri stars. If star
Ab were the source of the 7.38-day signal, the photometric amplitude
corrected for the light contribution of the other components of the
quadruple system (Table~\ref{tab:constraints}) would be in excess of
1.5 mag. This is a very rare occurrence among WTTS \citep[typical
amplitudes are about 0.1--0.2 mag; see, e.g.,][]{Stassun:99,
Grankin:08}.  On the other hand, if attributed to star Aa, the
intrinsic amplitude would be only 0.24 mag, which is fairly common.
Similarly for the 4.06-day signal, the intrinsic amplitudes we would
infer for stars Ba and Bb are 0.5 mag and 0.12 mag. Though not
unprecedented, the larger value is seen much less frequently than the
smaller one in photometric studies of young stars. In any event,
assignment of the observed signal to the primary or secondary in this
case is less important, as rotational synchronization is likely to
occur in both components given the short period and circular orbit, as
discussed earlier.

With our measurement of the projected rotational velocity of Aa from
Table~\ref{tab:constraints} and knowledge of its rotation period, we
may compute its mean density in a similar way as done for \lk\,B with
a slight modification of eq.(\ref{eq:densB}):
\begin{eqnarray}
\label{eq:densA}
\log\rho_{\rm Aa} &=& -1.8724 + \log P_{\rm orb} -3\log P_{\rm rot} + \nonumber \\
                  & & 2 \log(K_{\rm Aa}+K_{\rm Ab}) + \log K_{\rm Ab} - \nonumber \\
                  & & 3 \log(v_{\rm Aa}\sin i) + 1.5\log(1-e^2)
\end{eqnarray}
However, this still requires the assumption that the spin axis of star
Aa is aligned with the axis of the orbit ($i_{\rm rot} = i_{\rm orb}$),
which may or may not be true given the longer orbital period of the
binary and its young age.

\begin{figure}
\epsscale{1.1}
\vskip 10pt
\plotone{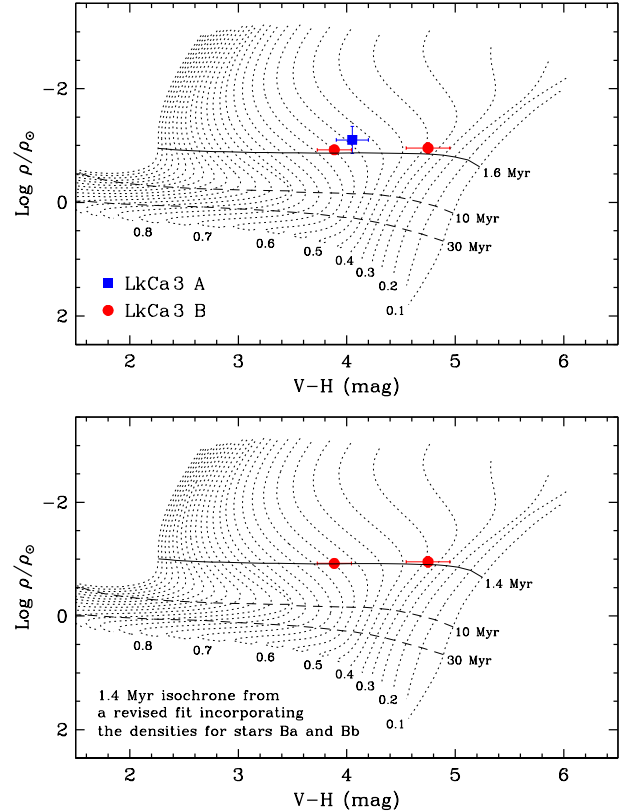}
\figcaption[]{\emph{Top:} Mean densities for stars Aa, Ba, and Bb
based on our $v \sin i$ measurements and adopted rotation periods (see
text), compared against the same PMS stellar evolution models from the
Dartmouth series shown in Figure~\ref{fig:models}. \emph{Bottom:} As
above, with the best-fit isochrone changed to 1.4 Myr from a new fit
incorporating the mean densities of stars Ba and
Bb.\label{fig:modelsdens}}
\end{figure}

\begin{figure*}
\epsscale{1.05}
\plottwo{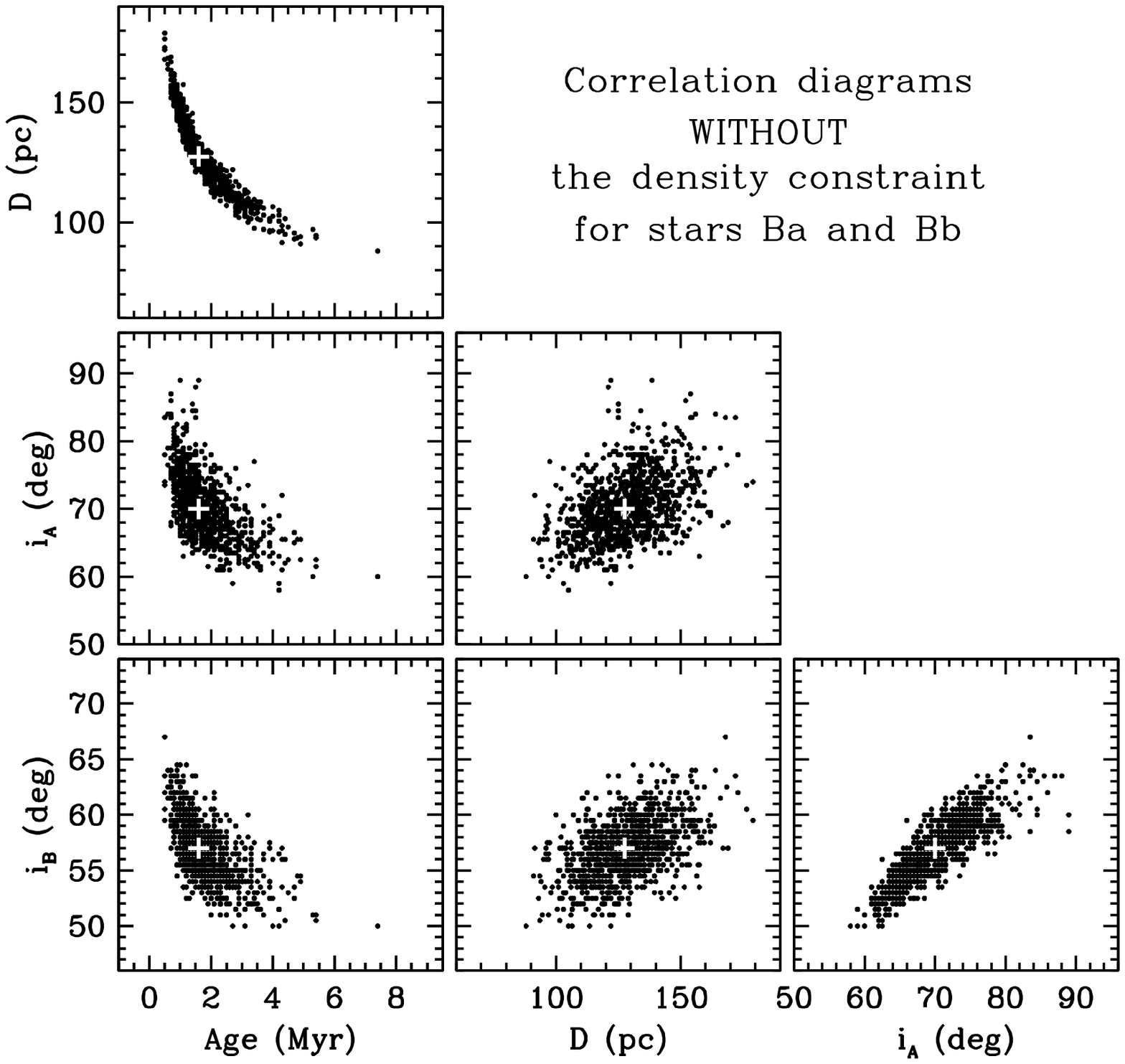}{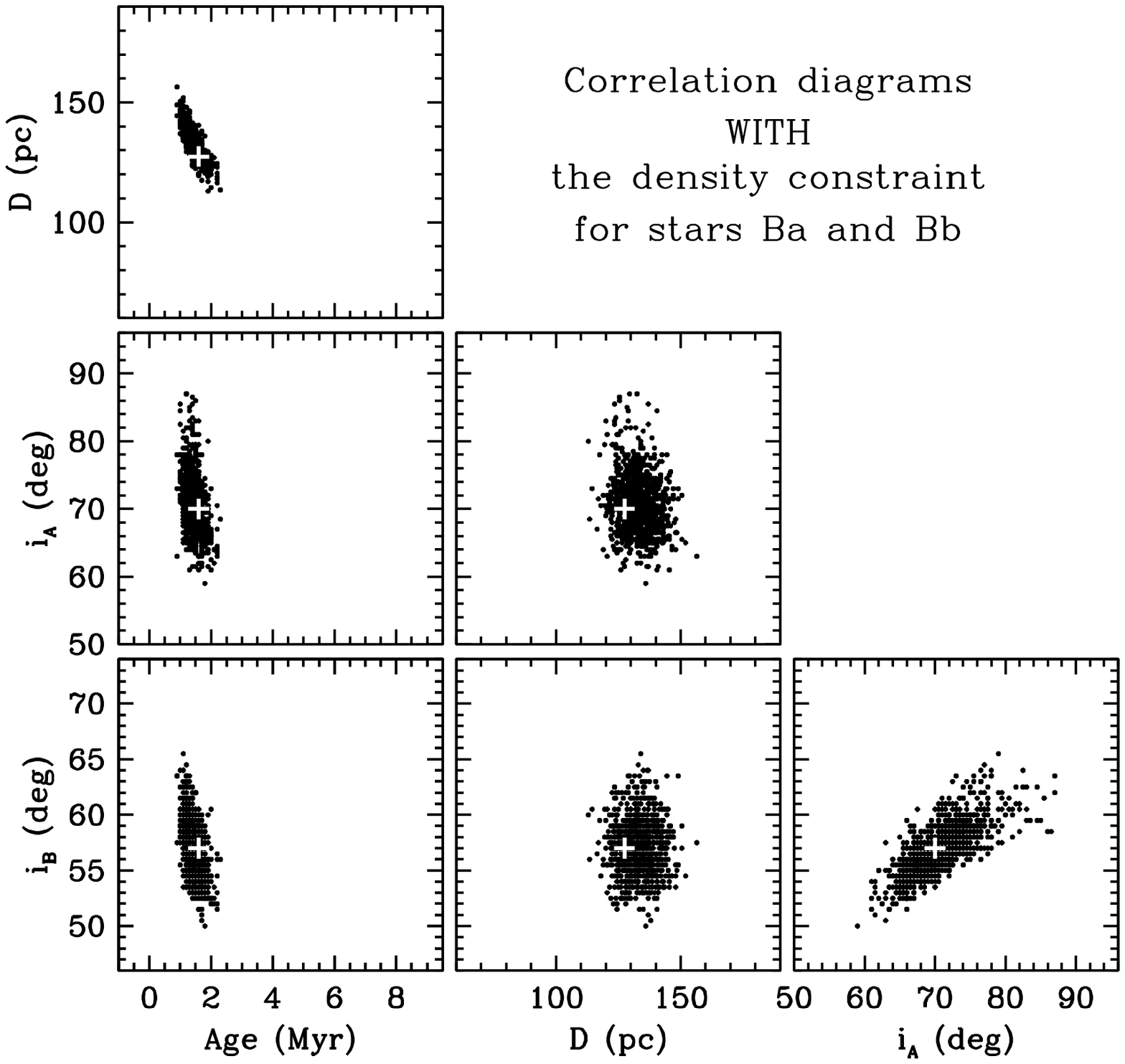}
\figcaption[]{\emph{Left:} Correlation diagrams between age, distance,
and the orbital inclination angles of \lk\,A and B resulting from our
Monte Carlo simulations to find the best fit between the observations
and the Dartmouth models. Crosses mark the best fit parameters. These
simulations did not make use of the constraint on the stellar
densities of stars Ba and Bb that comes from $v \sin i$.
\emph{Right:} The same diagrams from simulations now including the
density constraints. This largely removes the degeneracies between
various parameters, particularly age and distance.
\label{fig:correlations}}
\end{figure*}

The top panel of Figure~\ref{fig:modelsdens} shows that the mean
densities of stars Ba and Bb computed in this way, and also that of
Aa, seem to agree quite well with predictions from our best-fit
1.6-Myr isochrone, even though they were not used in that fit. It is
important to note that in our earlier model comparison there is a
certain amount of degeneracy between distance and age because
evolutionary tracks are mostly vertical in a diagram of magnitude
versus color (or temperature) such as Figure~\ref{fig:models}; a
similarly good agreement with the models can be obtained for a smaller
distance if one chooses an older age. On the other hand, the mean
stellar densities from eq.(\ref{eq:densB}) and eq.(\ref{eq:densA}) are
completely independent of distance, but strongly constrain the age,
and are therefore complementary to the brightness measurements. Since
the top panel of Figure~\ref{fig:modelsdens} suggests a somewhat
better fit may be had with a slightly younger age, we repeated the
comparison incorporating the densities of stars Ba and Bb as
additional constraints, though not that of star Aa because of
lingering doubts regarding its spin-orbit alignment. The new fit gives
an age of $1.4 \pm 0.2$ Myr, a distance of $D =
133.5_{-6.5}^{+6.0}$\,pc, and inclination angles for the binaries of
$i_{\rm A} = 70.5_{-3.5}^{+4.5}$ deg and $i_{\rm B} = 57.0 \pm 2.5$
deg, which are not very different from our previous results. This new
best-fit isochrone is shown with the density measurements in the
bottom panel of Figure~\ref{fig:modelsdens}.  The agreement with the
flux measurements is not significantly altered.  The uncertainties are
now smaller, especially for the distance and age, because use of the
densities suppresses the age-distance correlation to a significant
degree. This is illustrated graphically in
Figure~\ref{fig:correlations}. The masses of the stars do not change
appreciably, but the radii are now somewhat larger because of the
slightly younger age (approximately 1.63\,$R_{\sun}$ for stars Aa and
Ba, and 1.27\,$R_{\sun}$ for Ab and Bb).

\subsection{Exploring different distance estimates}
\label{sec:distance}

Given the multiple constraints on the models afforded by the
observations for \lk, we investigated the possibility that the data
might enable us to determine the distance to each binary
independently, and perhaps disprove our assumption that \lk\,A and B
are physically associated.  Because some of these constraints
including the flux ratios and combined magnitudes involve properties
of both binaries, it is not possible to carry out completely
independent solutions as the individual colors and magnitudes of the
binary components are strongly correlated. Instead, we performed a
joint solution involving all four stars as before, but we allowed the
distance as well as the age and inclination angle of each binary to
vary independently, for a total of six free parameters.  However,
since the mean stellar density constraint is applied only for stars Ba
and Bb, the parameters for \lk\,A are more poorly determined, and the
distances turn out to be indistinguishable within their formal errors:
$142_{-21}^{+22}$\,pc for \lk\,A, and $135.5_{-9.6}^{+7.5}$\,pc for
\lk\,B. The ages are $0.9_{-0.3}^{+0.8}$\,Myr and
$1.6_{-0.2}^{+0.4}$\,Myr, and the inclination angles $i_{\rm A} =
76.6_{-6.7}^{+17.4}$\,deg and $i_{\rm B} = 55.5_{-2.0}^{+2.6}$\,deg,
respectively. The parameters for \lk\,B are close to those reported
previously, as expected from the strength of the density constraint.

\section{Discussion}
\label{sec:discussion}

\begin{figure}
\epsscale{1.1}
\vskip 10pt
\plotone{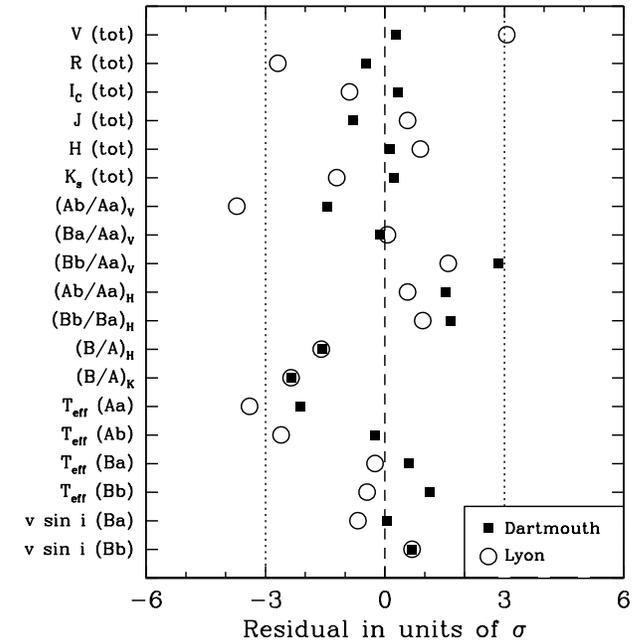}
\figcaption[]{$O-C$ residuals from the fits of the PMS isochrones to
the observations of \lk\ in Table~\ref{tab:constraints}, for both the
Dartmouth and Lyon models. The residuals are plotted in units of the
corresponding uncertainty of the measurement ($\sigma$) as given in
that table.  The dotted lines mark the $\pm$3$\sigma$ deviations.
\label{fig:residuals}}
\end{figure}

\lk\ is perhaps a unique case among PMS stars because of the many
observational constraints available for the comparison with stellar
evolution models, and the fact that we know these properties for four
stars in the same system that are presumably coeval. Beyond the
effective temperatures and relative fluxes in several passbands, which
are sometimes available for other PMS binaries, in \lk\ we have also
dynamical information in the form of minimum masses for all stars, and
the additional constraint from the mean densities for at least two of
the components. While the inferred age and particularly the distance
are in excellent agreement with expectations for the Taurus-Auriga
complex, these results are specific to the Dartmouth models of
\cite{Dotter:08}.  As a test we repeated the exercise with the earlier
but widely used series of Lyon models by \cite{Baraffe:98}, which
differ in the degree of convection assumed, among other physical
ingredients. While the Dartmouth models adopt a mixing length
parameter of $\alpha_{\rm ML} = 1.938$ in units of the pressure scale
height that yields a good fit to the solar properties, the publicly
available Lyon models for low-mass stars use $\alpha_{\rm ML} = 1.0$.
With these models we obtained a slightly older age of
$2.4_{-0.4}^{+0.1}$ Myr and a rather different distance of
$194_{-6}^{+12}$ pc for \lk\ that seems implausibly high for
Taurus-Auriga.  The quality of the fit is generally worse, giving a
$\chi^2$ value of $\sim$65 as opposed to $\sim$32.  A few of the
measurements deviate by more than 3$\sigma$, which was not the case
before, although there is no obvious pattern in the $O-C$ residuals
(see Figure~\ref{fig:residuals}) except that the predicted
temperatures seem consistently higher than observed.  The inclination
angles of the two binaries come out considerably smaller ($i_{\rm A} =
47.5 \pm 1.0$ deg and $i_{\rm B} = 41.0 \pm 1.0$ deg), resulting in
component masses that are essentially a factor of two larger than our
previous results. The primaries of the two binaries become
$\sim$1\,$M_{\sun}$ stars, which seems unreasonably high for an M
star.

\begin{figure}
\epsscale{1.1}
\vskip 10pt
\plotone{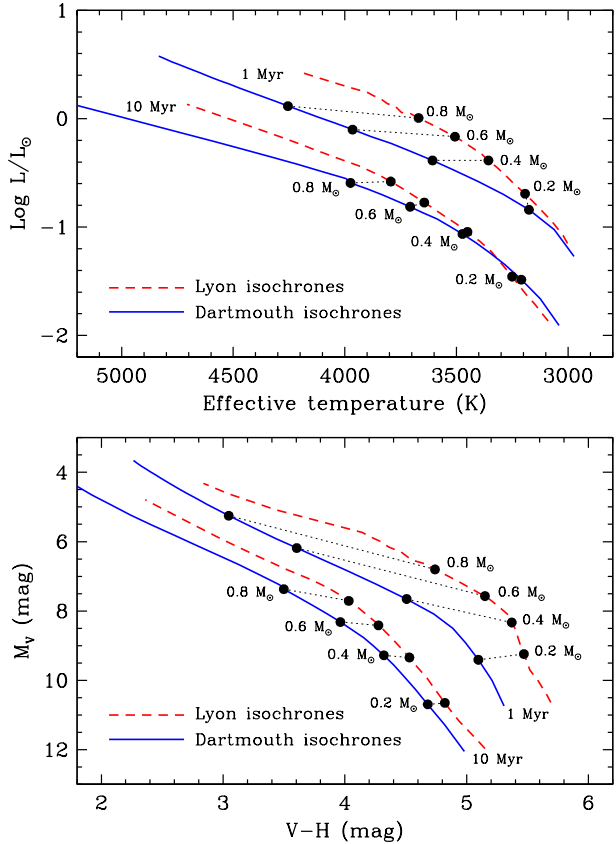}
\figcaption[]{Comparison of young solar-metallicity isochrones from
the Dartmouth series of \cite{Dotter:08} with similar models from the
Lyon series of \cite{Baraffe:98} that use $\alpha_{\rm ML} = 1.0$,
highlighting the significant differences.  \emph{Top:} Models in the
theoretical $\log L$ versus $T_{\rm eff}$ plane for ages of 1 and 10
Myr. Selected masses are indicated along each isochrone. {\it Bottom:}
Same isochrones as above, represented in the observational plane of
$M_V$ versus $V-H$ color.
\label{fig:comparemodels}}
\end{figure}

A more direct comparison between the Dartmouth and Lyon models is
illustrated in Figure~\ref{fig:comparemodels}, for two representative
solar-metallicity isochrones of 1\,Myr and 10\,Myr. The differences
are shown both in the theoretical plane of $\log L/L_{\sun}$ vs.\
$T_{\rm eff}$ and in the observational plane of $M_V$ vs.\ $V-H$
employed earlier in this work. Points with equal mass on both sets of
models are connected with dotted lines. Several features are
noteworthy. The Lyon models of \cite{Baraffe:98} appear systematically
higher in the diagrams, particularly at 1\,Myr, explaining the larger
distance obtained. While the bolometric luminosities are similar in
both models at a given mass, the temperatures are higher in the
Dartmouth series, increasingly so for larger masses. In the
observational plane there are large differences in color of nearly 2
mag for $M = 0.8\,M_{\sun}$, which are partly attributable to the
temperature differences just mentioned, and perhaps also to the
color/temperature transformations used in each model. This is the
reason for the much larger masses we obtain for \lk\ when using the
\cite{Baraffe:98} models.
All of these differences are seen to be larger at younger ages, which
is where the models are most sensitive to initial conditions, and also
where the calculations are known to be less reliable \citep[see,
e.g.,][]{Tout:99, Siess:01, Baraffe:02}. 
We note also that neither set of models incorporates the effects of
magnetic fields, which are ubiquitous among PMS stars and have been
shown to have a significant impact on theoretical predictions for
young stars \citep[e.g.,][]{DAntona:00}. The same has been found more
generally for late-type active main-sequence stars
\citep[e.g.,][]{Mullan:01, Chabrier:07, Feiden:12}.

The popularity of the Lyon models with $\alpha_{\rm ML} = 1.0$ for PMS
stars and the relatively infrequent use the Dartmouth models have had
so far in this area of research motivate us to examine these
discrepancies in more detail below, for the impact they may have on
our knowledge of young stars.

\subsection{The discrepancies with the Lyon models}
\label{sec:lyon}

The poor performance of the Lyon models of \cite{Baraffe:98} in the
case of \lk\ may seem surprising given that they have been employed in
numerous studies since their appearance, and have even been used to
help define the temperature scale for late-type PMS stars (see
below). On relatively few occasions have difficulties with these
models been documented in the young-star literature \citep[for
examples see][]{Luhman:98, Hillenbrand:04}.
This may have to do with the nature and large number of observational
constraints for \lk, in contrast with the typical situation for young
stars.
A somewhat similar case to \lk\ in terms of the strength of the
observational constraints is that of GG\,Tau, which is another
hierarchical quadruple PMS system comprised of two visual binaries in
the Taurus-Auriga star-forming region. This system is often regarded
as a benchmark for testing evolutionary calculations for young stars,
and the observations for GG\,Tau are generally considered to support
the validity of the Lyon models, in apparent contradiction with our
findings for \lk.

For GG\,Tau the individual spectral types (K7+M0 and M5+M7) and
luminosities (for an assumed distance of 140\,pc) are known for all
four components from the work of \cite{White:99}. Additionally, the
total mass of one of the binaries (GG\,Tau\,A) has been estimated from
measurements of the orbital velocities of the circumbinary disk
\citep[$M_{\rm Aa+Ab} = 1.28 \pm 0.07\,M_{\sun}$;][]{Dutrey:94,
Guilloteau:99}. As in \lk, the four stars in GG\,Tau form a
``mini-cluster'' of sorts (on the assumption that they are coeval),
which means they provide an unusually strong constraint on stellar
evolution theory because a successful model must be able to reproduce
the properties of all four components at a single age. This was
exploited by \cite{White:99} to test the Lyon models, among others,
and in the process to establish a conversion between spectral types
and effective temperatures for late-type PMS stars, which has been a
persistent source of uncertainty in the field.

The authors showed that the solar-metallicity Lyon models with
$\alpha_{\rm ML} = 1.0$, which are the same ones we used for \lk,
provide a good match to the effective temperatures and luminosities of
all four stars in GG\,Tau (see their Fig.\,6), but yield a predicted
total mass for GG\,Tau\,A of $2.00 \pm 0.17\,M_{\sun}$ that is
considerably larger than observed. Thus, these models fail as in the
case of \lk.  The temperatures in this analysis were converted from
spectral types, but were allowed some freedom for the cooler stars Ba
and Bb in order to produce the best fit, which resulted in an implied
temperature scale for the SpT/$T_{\rm eff}$ conversion that is
intermediate between dwarfs and giants. The two scales converge for
the hotter stars Aa and Ab, which were not adjusted
separately. \cite{White:99} also tested Lyon models computed with a
higher mixing length parameter of $\alpha_{\rm ML} = 1.9$ appropriate
for the Sun. These were found to produce a similarly good agreement
with the temperatures (readjusted for Ba and Bb) and luminosities, and
a lower mass for GG\,Tau\,A of $1.46 \pm 0.10\,M_{\sun}$ that is in
better accord with the dynamically measured value. Very similar
conclusions were reached by \cite{Luhman:99} for the $\alpha_{\rm ML}
= 1.9$ models from Lyon, apparently lending support to their accuracy.

We point out, however, that the $\alpha_{\rm ML} = 1.9$ models
employed by the above authors are actually a ``hybrid'', since
published tabulations from the Lyon group with this mixing length
parameter only reach down to 0.6\,$M_{\sun}$. Below this mass the
authors reverted to the \cite{Baraffe:98} models with $\alpha_{\rm ML}
= 1.0$, on the assumption that variations in $\alpha_{\rm ML}$ are
inconsequential under 0.6\,$M_{\sun}$ \citep{Chabrier:97,
Baraffe:98}. As it turns out, 1\,Myr models for the two values of
$\alpha_{\rm ML}$ have about the same luminosity at this mass, but
temperatures that differ significantly by some 230\,K, with the
$\alpha_{\rm ML} = 1.9$ calculations being hotter \citep[see
also][]{Weinberger:13}.  Both \cite{White:99} and \cite{Luhman:99}
were then forced to artificially connect the $\alpha_{\rm ML} = 1.0$
isochrones below 0.6\,$M_{\sun}$ in some smooth fashion with the
$\alpha_{\rm ML} = 1.9$ isochrones at some mass above 0.6\,$M_{\sun}$
in order to bridge the discontinuity, which results in a kink at
0.6\,$M_{\sun}$ and an unsmooth and likely unphysical change in
stellar properties across the boundary.

This situation is less than ideal because the $\alpha_{\rm ML} = 1.9$
models were computed with a helium abundance $Y$ appropriate for the
Sun that is not the same as that assumed for the $\alpha_{\rm ML} =
1.0$ models. A larger $Y$ for these young ages (as in the $\alpha_{\rm
ML} = 1.9$ models) typically leads to a lower luminosity at a given
temperature, with some dependence on mass, so in a strict sense
joining the isochrones is inconsistent.  Furthermore, the assumption
that the models become insensitive to the mixing length parameter
below about 0.6\,$M_{\sun}$ is not necessarily valid at early ages for
stellar models with low surface gravities and cool temperatures
\citep{Baraffe:02}, and may only hold true for much lower masses
\citep[see also][]{Burrows:89, DAntona:94, Luhman:98}. Indeed, a
comparison of published Lyon models with $\alpha_{\rm ML} = 1.9$ and
$\alpha_{\rm ML} = 1.0$ shows that the difference in effective
temperature at a fixed mass of 0.6\,$M_{\sun}$ rises considerably
toward younger ages, from about 60\,K at 10\,Myr, to 120\,K at 3\,Myr,
and 230\,K at 1\,Myr, which is the age regime relevant for GG\,Tau.

This undermines the conclusion that the Lyon models for ``$\alpha_{\rm
ML} = 1.9$'' used by \cite{White:99} and \cite{Luhman:99} agree with
the measurements for GG\,Tau, as the comparison may not be very
meaningful.

\begin{figure}
\epsscale{1.15}
\vskip 10pt
\plotone{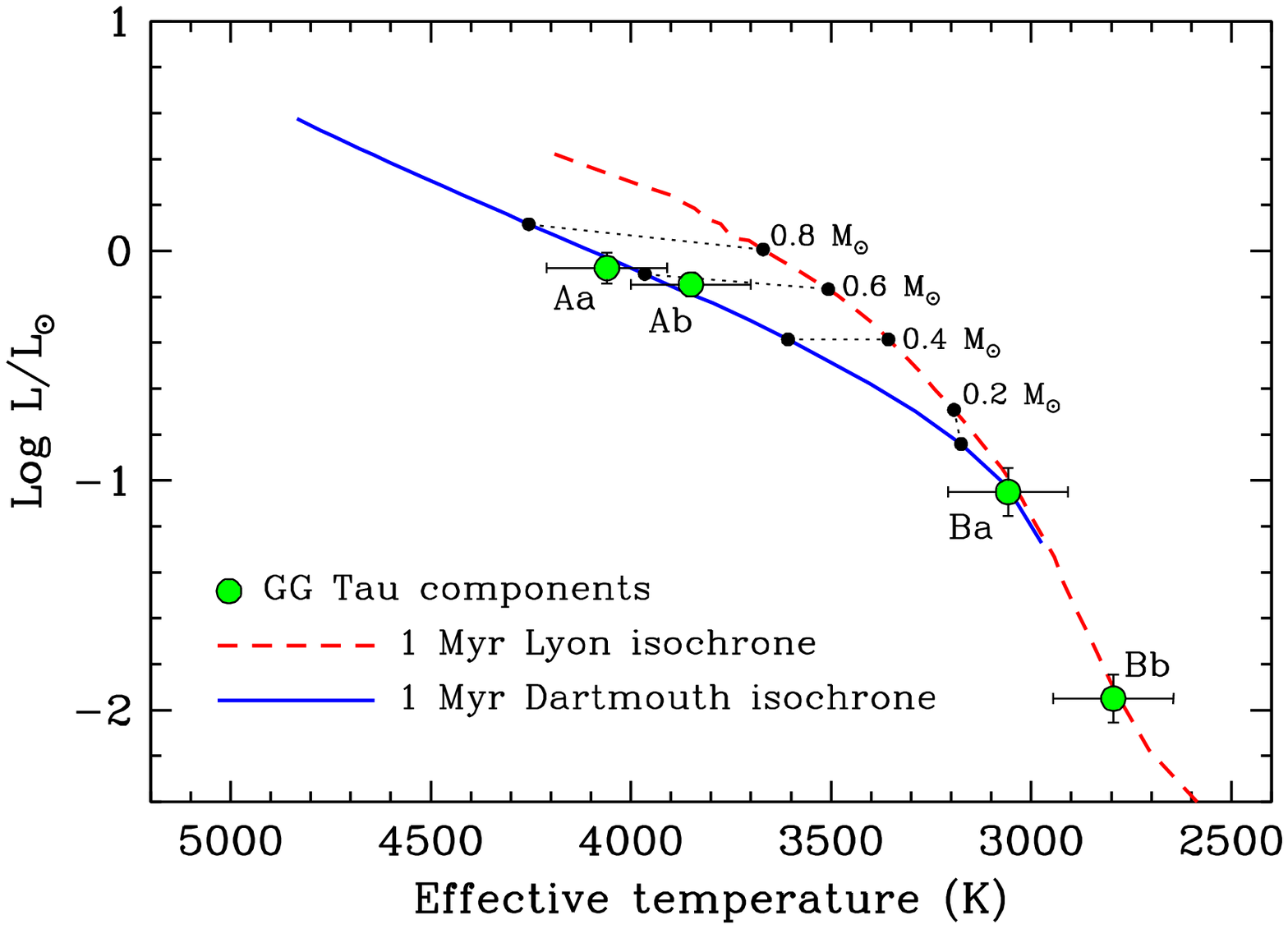}
\figcaption[]{Bolometric luminosities and effective temperatures for
  the components of the quadruple PMS system GG\,Tau \citep{White:99,
    Luhman:99} compared against 1-Myr solar-metallicity isochrones
  from the Dartmouth series of \cite{Dotter:08} and the Lyon series of
  \cite{Baraffe:98} for $\alpha_{\rm ML} = 1.0$, as displayed in the
  top panel of Figure~\ref{fig:comparemodels}. The Lyon model extends
  to lower masses than available in the Dartmouth calculations.
  Reference masses (small circles) are labeled along both models and
  are connected with dotted lines.\label{fig:ggtau}}
\end{figure}

Given that the Dartmouth models seem to work well for \lk, it is of
interest to see how they fare when tested against the observed
properties of GG\,Tau.  This comparison is shown in
Figure~\ref{fig:ggtau}, for an age of 1\,Myr.  The spectral types and
luminosities adopted for stars Aa and Ab are those reported by
\cite{White:99}, and for stars Ba and Bb we used the revised
determinations by \cite{Luhman:99}.  Spectral types were converted to
effective temperatures with the intermediate scale of \cite{Luhman:99}
for Ba and Bb, and the dwarf scale for Aa and Ab since the difference
with the giant scale is insignificant at these earlier spectral
types.\footnote{\cite{Luhman:03a} reported a minor revision of the
temperature scale that has a negligible effect on our conclusions, as
it implies only a 10\,K reduction in $T_{\rm eff}$ for star Bb.} The
temperatures were assigned a conservative uncertainty of 150\,K,
corresponding approximately to a spectral type error of one subtype.

In Figure~\ref{fig:ggtau} star Bb is outside the range of masses
covered by the Dartmouth models, but there is excellent agreement for
the other three components of GG\,Tau.  The dynamical mass estimate
for GG\,Tau\,A is also well reproduced by these models: the sum of the
inferred masses of stars Aa and Ab is $1.19 \pm 0.06\,M_{\sun}$, to be
compared with the measured value of $1.28 \pm 0.07\,M_{\sun}$. The
individual masses we derive from the Dartmouth models are $M_{\rm Aa}
= 0.63 \pm 0.05\,M_{\sun}$, $M_{\rm Ab} = 0.56 \pm 0.04\,M_{\sun}$,
and $M_{\rm Ba} = 0.15 \pm 0.02\,M_{\sun}$.

We note that the SpT/$T_{\rm eff}$ conversion of \cite{Luhman:99} used
in this comparison was established largely using GG\,Tau itself
(specifically, the Ba and Bb components) along with the hybrid
``$\alpha_{\rm ML}$ = 1.9'' models from Lyon. Using this to test the
Dartmouth models may therefore seem to be inconsistent. However, as
seen in Figure~\ref{fig:ggtau}, the Lyon and Dartmouth isochrones
converge near star Ba (slightly under 0.2\,$M_{\sun}$), so the fact
that the Dartmouth model is able to match the temperature of this star
is not surprising. Whether it would also be able to reproduce the
temperature of the Bb component using the same SpT/$T_{\rm eff}$
conversion, if the calculations were extended to lower masses, remains
to be seen.

We conclude from the above that, as in the case of \lk, the available
constraints for the quadruple system GG\,Tau favor the PMS Dartmouth
models over those from Lyon with $\alpha_{\rm ML} = 1.0$, at least at
these young ages. It is quite possible, however, that the Lyon models
with the higher value of $\alpha_{\rm ML} = 1.9$, if they were
available for masses below 0.6\,$M_{\sun}$, might also perform well,
although this cannot be verified at the moment. A second conclusion we
may draw is that the Dartmouth models do not appear to be inconsistent
with the SpT/$T_{\rm eff}$ prescription advocated by \cite{Luhman:99},
in which a dwarf-like temperature scale is adopted for young stars
earlier than M0 and an intermediate scale for later-type stars,
although further observational constraints between M1 and M5 are
required to confirm this.

Based on the examples of GG\,Tau and \lk\ we would expect that in
other cases the use of the Dartmouth models instead of the
$\alpha_{\rm ML} = 1.0$ Lyon models for PMS stars would result in
generally smaller inferred masses, and slightly younger ages, although
this will depend to some extent on the constraints available.

\section{Final remarks}
\label{sec:finalremarks}

The quadruple system \lk\ presented here provides an unusually
stringent test of PMS evolution models that appears even stronger than
the classical case of GG\,Tau. In both systems what makes the
constraint on theory so strong is the dynamical mass information, and
in \lk\ the additional estimates of the mean stellar densities for two
of the components.  While young clusters, associations, and larger
star-forming regions such as IC\,348, $\epsilon$\,Cha, $\eta$\,Cha,
Chamaeleon\,I, Taurus \citep{Luhman:99, Luhman:04a, Luhman:04b,
Luhman:03a, Luhman:03b} and many others do provide some ability to
test models from the fact that they are coeval populations, and can
also help to establish the SpT/$T_{\rm eff}$ relation for PMS stars,
the constraints they afford are not as tight because of the often
large scatter of the measurements in the H-R diagram, and the lack of
information on mass.

The Dartmouth models of \cite{Dotter:08}, with their consistent
physics across the entire mass range, are strongly favored by the
observations of \lk\ and GG\,Tau over the Lyon models with
$\alpha_{\rm ML} = 1.0$. In retrospect, the performance of the latter
models when confronted with the challenge of multiple complementary
constraints for the same stellar system is not entirely surprising, as
there is no obvious physical reason why the mixing length parameter
should have precisely that value.
The present study has shown that the Dartmouth models are also
preferred over the hybrid Lyon models with ``$\alpha_{\rm ML} = 1.9$''
considered by \cite{White:99}, \cite{Luhman:99}, and others, which
\emph{do not} use consistent physics.

Other than the present work, relatively few tests of the Dartmouth
models for pre-main sequence stars have been reported in the
literature, but are highly desirable to strengthen our confidence in
the predictive power of these calculations. Binary and multiple
systems are especially suited for this, and other information for
these or other young objects can be extremely helpful such as
trigonometric parallaxes (which the GAIA space mission is expected to
supply in the near future), angular diameters, and chemical analyses
including determinations of the lithium abundance.  For example,
although strong \ion{Li}{1} $\lambda$6707 in absorption is present in
\lk, the multiplicity of the system has so far made a quantitative
interpretation of those measurements impossible. The recent
high-resolution spectroscopic study by \cite{Nguyen:12} did not report
Li measurements, but a graphical representation of the Li region for
their four spectra shows several components which we can now identify
from our orbital solutions. Stars Aa, Ba, and Bb all exhibit the line
in absorption, though we see no sign of star Ab, the faintest one of
the four. With knowledge of the flux ratios from our work, it should
now be possible to re-analyze the original \cite{Nguyen:12} spectra
and derive Li abundances for at least three of the stars, properly
corrected for the light contribution from the other components. This
could serve as an important additional constraint on stellar evolution
models.

\acknowledgements

We thank the anonymous referee for interesting suggestions, as well as
  I.\ Baraffe and R.\ White for helpful exchanges about stellar
  evolution models.  The spectroscopic observations at the CfA were
  obtained with the able assistance of
P.\ Berlind,
R.\ Davis,
L.\ Hartmann,
E.\ Horine,
A.\ Milone, and
J.\ Peters.
We are grateful to R.\ Davis for maintaining the CfA echelle database
over the years.  We also thank C.\ Beichman for the opportunity to
obtain an additional NIRSPEC spectrum on the night of UT 2010 November
22.  We thank the staff at Keck Observatory for their superb support.
Data presented herein were obtained at the W.\ M.\ Keck Observatory
from telescope time allocated to the National Aeronautics and Space
Administration through the agency's scientific partnership with the
California Institute of Technology and the University of California.
LP acknowledges support from NASA Keck PI Data Award administered by
NExScI (semesters 2003B and 2004A).  Keck telescope time was also
granted by NOAO, through the Telescope System Instrumentation Program
(TSIP). TSIP was funded by the NSF. The Observatory was made possible
by the generous financial support of the W.\ M.\ Keck Foundation.  We
recognize the Hawaiian community for the opportunity to conduct these
observations from the summit of Mauna Kea. This work was partially
supported by NSF grants AST-1007992 to GT and AST-1009136 to LP. The
research has made use of NASA's Astrophysics Data System Abstract
Service, and of the SIMBAD and VizieR databases, operated at the CDS,
Strasbourg, France.


\end{document}